\documentclass[reprint,superscriptaddress,showpacs,nofootinbib,aps,prb]{revtex4-1}
\usepackage{amssymb}
\usepackage{amsmath}
\usepackage{graphicx}
\usepackage{epstopdf}
\usepackage{dcolumn}
\usepackage{bm}
\usepackage[utf8]{inputenc}
\usepackage{braket}

\begin{document}

\title{Fragmented-condensate solid of dipolar excitons}
\author{S. V. Andreev}
\email[Electronic adress: ]{Serguey.Andreev@gmail.com}
\affiliation{ITMO University, St. Petersburg 197101, Russia}
\affiliation{LPTMS, CNRS, University Paris Sud, UMR8626, 91405 Orsay, France}
\affiliation{University of Bordeaux, LOMA UMR-CNRS 5798, F-33405 Talence Cedex, France}
\date{\today}

\begin{abstract}
We discuss a possible link between the recently observed macroscopic ordering of ultracold dipolar excitons (MOES) and the phenomenon of supersolidity. In the dilute limit we predict a stable supersolid state for a quasi-one-dimensional system of bosonic dipoles characterized by two- and three-body contact repulsion. We phenomenologically extend our theory to the strongly-correlated regime and find a critical value of the contact interaction parameter at which the supersolid exhibits a quantum phase transition to a fragmented state. The wavelength of the \textit{fragmented-condensate solid} is defined by the balance between the quantum pressure and the entropy due to fluctuations of the relative phases between the fragments. Our model appears to be in good agreement with the relevant experimental data, including the very recent results on commensurability effect and wavelength of the MOES.    
\end{abstract}

\pacs{Valid PACS appear here}
\maketitle

\preprint{APS/123-QED}

\section{Introduction}

Quest for supersolids ("coherent crystals") in Bose-Einstein condensed systems is a subject with long history dating back to the general argument by Gross\cite{Gross} in 1957 and the 1971 theoretical proposal by Kirzhnits and Nepomnyashchii \cite{Kirzhnits}. Provided that the effective two-body interaction between the particles is designed in such a way, that it has sufficiently large negative Fourier components in the vicinity of some finite momentum transfer $\bm k_0$, the ground state of the system can exhibit crystalline order. Particular interest represent long-wavelength structures, with unit cells containing macroscopically large amount of particles. Despite numerous theoretical proposals \cite{Nepomnyashchii, Brazovskii, Pitaevskii, Suto, Pomeau, Meystre, Rica, Henkel, Li, Cinti, Saccani, Nozieres, Chen, Josserand, DiluteSupersolid}, no convincing evidence of existence of such structures in nature has been reported.

Following significant advances in creation of ultracold polar molecules \cite{Molecules}, a possibility of a supersolid with dipolar gases tightly confined to two dimensions (2D) and having dipole moments oriented perpendicular to the plane of their translational motion has been theoretically considered \cite{Li,Meystre}. These studies were triggered by prediction of the helium-like roton-maxon instability \cite{Santos} and are now focused on the properties of possible quantum phases and transitions between them. A model of a stable supersolid of dipolar bosons has been proposed in the dilute limit at absolute zero temperature \cite{DiluteSupersolid}. The $T=0$ requirement is indispensable in reduced dimensionality to guarantee the extension of the phase coherence over multiple periods of the structure. Thus conceived state of matter combines the crystalline order with the properties of a superfluid: it is commensurate, but its period can be tuned by varying the density or the velocity of the system.

Recently, quantum degenerate gases of dipolar excitons have been realized in semiconductor quantum wells\cite{ColdExcitons}. Large binding energy of 2D excitons makes it possible to create dense and strongly-correlated ensembles\cite{dense}, whereas the small effective mass raises the ultracold limit temperatures to the range of few degrees Kelvin. Powerful methods of semiconductor optical spectroscopy can be applied to excitonic gases in order to study their properties\cite{Ivchenko}. For instance, photoluminescense (PL) measurements provide access to the exciton density\cite{Bieker}, spin\cite{Kavokin} and energy distribution\cite{Pfeiffer}. Shift-interferometry of the exciton PL offers a unique possibility for systematic investigation of coherence properties of a gas without perturbing its (quasi-)equilibrium configuration (this is in contrast with time-of-flight \cite{Andrews}, Bragg spectroscopy \cite{Ketterle} and "momentum focusing" \cite{Walraven} techniques employed for atomic systems).

One of the most intriguing features observed in the experiments on ultracold excitons is the macroscopically ordered exciton state (MOES) \cite{Butov2002}. Below some temperature (up to 4 K) the external ring in the PL pattern of indirect excitons fragments into a chain of regularly spaced aggregates ("beads") having macroscopic sizes. Local exciton energy follows the density distribution, so that the aggregates are characterized by strong repulsive interaction \cite{Repulsive}. The periodical density modulation is accompanied by a buildup of the off-diagonal long range order (ODLRO) at the edges of each bead. The coherence of the PL collected from the cores of the beads remains partially suppressed even at the lowest temperatures achieved in the experiment. The coherence length at the edges is comparable with the size of one bead and much less than the circumference of the ring\cite{High2012}. Along with the emergence of extended coherence and spatial periodicity, the dependence of the exciton energy on temperature changes derivative \cite{Repulsive}. The corresponding critical point $T_c$ is remarkably robust to disorder: independent segments of the ring isolated by defects (current filaments) and having strongly different lengths and widths fragment at the same temperature.  

Several theoretical models were proposed to describe the MOES \cite{Yang, Paraskevov, Levitov, Sugakov, Liu, Wilkes}. An explanation of the aforementioned experimental facts starting from a single principle was given in a series of papers \cite{Andreev1, Andreev2, Andreev3}. The microscopic mechanism underlying the phenomenon as proposed in this latter model is reminiscent of the roton instability in dipolar superfluids. It is clear, however, that the MOES cannot be regarded as a conventional supersolid at least because it has a very different coherence regime. An attempt to put the physics of the exciton density wave in the context of supersolidity in dipolar gases is the subject of this paper. Our main conclusion is that the MOES may be described as a \textit{fragmented-condensate solid}: a supersolid where the global coherence is destroyed by the quantum fluctuations of the relative phases between the adjacent lattice sites. This picture proves to be consistent not only with the earlier experimental results mentioned above, but is further supported by the recently observed commensurability of the MOES and dependence of its wavelength on the exciton density\cite{Commensurability}.

The paper is organized as follows. In Sec. II we present our theory. We introduce a minimal model describing a quasi-one-dimensional gas of dipolar excitons (which corresponds to a segment of the ring in the experiments). We also point out a possible realization of such model with ultra cold polar molecules. By using a mean field approach we discuss emergence of supersolidity in the system for two different regimes: a one-dimensional mean field and a 2D Thomas-Fermi cigar. We suggest that a stable dipolar supersolid can be described as a chain of self-trapped Bose-Einstein condensates. For sufficiently strong repulsive interaction between the particles the chain undergoes a quantum phase transition to a number-squeezed fragmented configuration, akin to the fragmented BEC in an optical lattice\cite{Kasevich}. We estimate the critical value of the relevant parameter and obtain an expression for the wavelength $\lambda$ of the fragmented-condensate solid. All results of the theory have transparent analytical form. 

In Sec. III we compare our results with the experiments. Previously, we have already demonstrated consistency of our formula for the MOES wavelength in the regime where the ring is allowed to expand when changing the laser excitation power (fixed gate voltage)\cite{Andreev1}. Here we use the same equation to fit the recently observed dependence of $\lambda$ on the gate voltage in the regime where the ring radius is maintained fixed. In this latter case $\lambda$ was found to increase when increasing the exciton density (the gate voltage)\cite{Commensurability}. This is in contrast to what would be expected for conventional supersolids. Our formula captures this experimental result as well. We conclude Sec. III by explaining the observed commensurability of the exciton density wave. 

In Sec. IV we list main results and conclusions. Being an effective field theory, our spinless model can be obtained from more realistic multi-component systems by tracing out the corresponding degree of freedom. In some cases, this may offer one a possibility to independently tune the short- and long-range parts of the effective interaction. We point out important examples of such systems in excitons and polar molecules. The underlying principle is a 2D analogy of the Feshbach resonance in atomic gases\cite{Andreev4}, and it should allow one to test the predictions of the theory in a controllable fashion. 

\section{Theoretical model}

We consider a spinless gas of purely 2D bosonic particles having dipole moments oriented perpendicularly to the plane of their center-of-mass translational motion ($xy$ plane). In order to model a segment of the ring we introduce harmonic confinement in one direction ($y$ axis). The Hamiltonian of the system reads
\begin{equation}
\label{SpinlessHamiltonian}
\begin{split}
\hat H=&\int\hat\Psi^\dagger(\bm \rho)\left(-\frac{\hbar^2}{2m}\Delta+\frac{m\omega_y^2y^2}{2}\right)\hat\Psi(\bm \rho) d\bm \rho\\
&+\frac{1}{2}\int\hat\Psi^\dagger(\bm \rho)\hat\Psi^\dagger(\bm \rho')V(\bm \rho-\bm \rho')\hat\Psi(\bm \rho')\hat\Psi(\bm \rho)d\bm \rho d\bm \rho',
\end{split}
\end{equation}
where $\bm\rho=(x,y)$, $m$ is the particle mass and $\omega_y$ is the confinement oscillator frequency. For our purposes it will be convenient to recast the particle field operator in the form
\begin{equation}
\label{ansatz}
\hat\Psi(x,y)=\hat\psi(x)\frac{\varphi(y)}{\sqrt{a_y}},
\end{equation}
and to expand its second-quantized part $\hat\psi(x)$ in terms of the plane waves
\begin{equation}
\label{planewave}
\hat\psi(x)=\frac{1}{\sqrt{L}}\sum_{k_x} \hat c_{k_x}e^{ik_x x},
\end{equation}
with $k_x=\{0,\pm 2\pi/L,\pm 4\pi/L,...\}$. Here $a_y=(\hbar/m\omega_y)^{1/2}$ and $L$ is the length of the segment. At distances much larger than the transverse size of the system the two-body interaction potential $V(\bm \rho-\bm \rho')$ behaves as
\begin{equation}
\label{tail}
V(\bm \rho-\bm \rho')\approx V_\ast(x-x')=\frac{\hbar^2}{m}\frac{x_\ast}{\lvert x-x'\rvert^3},
\end{equation}
where $x_\ast=m\mathrm{e}^2d^2/4\pi\hbar^2\epsilon\epsilon_0$ is the characteristic dipole-dipole distance. In what follows we shall assume $ a_y\gg x_\ast$.

At zero temperature one can expect a range of parameters (to be specified below) in which the system can be described by the uniform (along $x$) order parameter
$\Psi(y)=\sqrt{n_1}\varphi(y)/\sqrt{a_y}$
with $n_1=N/L$ being the 1D density. In the lowest-order approximation 
\begin{equation}
\label{Contact}
V(\bm \rho-\bm \rho')=V_0\delta(\bm\rho-\bm\rho'),
\end{equation}
where $V_0$ is the $\bm q=0$ value of the momentum-dependent pseudopotential\cite{Pitaevskii}, and the function $\varphi(\tilde y)$ can be found by solving the Gross-Pitaevskii equation
\begin{equation}
\label{GP}
\left (-\frac{\partial^2}{\partial\tilde y^2}+\frac{\tilde y^2}{2}+\frac{mV_0}{\hbar^2} n_1 a_y \varphi^2\right) \varphi =\frac{\mu}{\hbar\omega_y}\varphi,
\end{equation}
where $\tilde y=y/a_y$. The chemical potential $\mu$ is defined by the normalization condition
\begin{equation}
\label{norm}
\int \varphi^2 (\tilde y) d\tilde y=1.
\end{equation}
The dimensionless form \eqref{GP} allows one to identify two important regimes of the mean field approximation. The first one, hereafter called the \textit{1D mean field}, corresponds to the range of densities where
\begin{equation}
\label{MF}
\frac{mV_0}{\hbar^2}\ll n_1 a_y \ll \frac{\hbar^2}{mV_0},
\end{equation}
and interactions are weak ($mV_0/\hbar^2<1$). In this case the solution of \eqref{GP} approaches the Gaussian ground state of the transverse harmonic trap. The chemical potential is given by
\begin{equation}
\label{muMF}
\mu=\frac{\hbar\omega_y}{2}+V_0^{1D}n_1,
\end{equation}
where we have introduced the 1D coupling constant
\begin{equation}
\label{V01D}
 V_0^{1D}=V_0/\sqrt{2\pi} a_y.
\end{equation}
Strictly speaking, there is no true condensate in 1D even at zero temperature. The one-body density matrix behaves as\cite{Haldane}
\begin{equation}
\label{onebodydensity}
n^{(1)}(s)\propto (\xi/s)^{\nu}
\end{equation}
at large distances $s=|x-x'|$, with $\xi=\hbar/\sqrt{2mV_0^{1D}n_1}$ being the healing length and
\begin{equation}
\label{nu}
\nu=mc_1/2\pi \hbar n_1.
\end{equation}
Thus, the ODLRO does not extend to infinity as in higher dimensions. However, in the limit \eqref{MF} one has $\nu\ll 1$ [this can be easily seen by substituting Eq. \eqref{c1} for $c_1$ into \eqref{nu}], so that the order extends up to macroscopic distances much larger than $\xi$ [and, hence, spreads over a large amount of the lattice sites in the dilute supersolid phase, see Subsection B]. This justifies the use of the mean-field approach\cite{Popov}.         

In the opposite limit, where
\begin{equation}
\label{TF}
n_1 a_y \gg \frac{\hbar^2}{mV_0},
\end{equation}
one enters the transverse Thomas-Fermi regime, which we shall refer to as \textit{the 2D cigar}. In this regime the system locally retains its 2D origin. By imposing the normalization condition \eqref{norm} on $f$ one obtains
\begin{equation}
\label{muTF}
\mu=\frac{\hbar\omega_y}{2}\left(\frac{3}{2}\frac{mV_0}{\hbar^2}n_1 a_y\right)^{2/3}.
\end{equation}
By using the relationship
\begin{equation}
\label{TDmu}
mc_1^2=n_1\frac{\partial\mu}{\partial n_1}
\end{equation}
and Eq. \eqref{nu}, one finds
\begin{equation}
\nu=\left(\frac{3 m V_0}{2 \hbar^2 n_1^2 a_y^2}\right)^{1/3}
\end{equation}
for the power of the characteristic decay law \eqref{onebodydensity}. Again, one has $\nu\ll 1$, provided that the condition \eqref{TF} is satisfied. The Thomas-Fermi half-width of the cigar is defined in the usual way as a classical turning point and it reads
\begin{equation}
\label{RTF}
R_c=a_y\left(\frac{3}{2}\frac{mV_0}{\hbar^2}n_1 a_y\right)^{1/3}.
\end{equation}
The structure of Eq. \eqref{muTF} implies that the thermodynamic limit for such quasi-1D system can be achieved by letting simultaneously $N\rightarrow\infty$, $L\rightarrow\infty$ and $\omega_y\rightarrow 0$, while keeping fixed the combination $N\omega_y/L$. According to \eqref{RTF} the width of the cigar in this limit becomes increasingly large. Throughout the paper, however, we shall always assume $R_c\ll L$, so that the system looks one-dimensional from the geometrical point of view.

For strong interactions ($mV_0/\hbar^2>1$) the Gross-Pitaevskii picture discussed above becomes inadequate. In this case one can distinguish between a Tonks-Girardeau gas of impenetrable bosons\cite{Tonks} (dilute limit, $n_1a_y<\hbar^2/mV_0$) and a dense strongly correlated quasi-1D superfluid ($n_1a_y\gg1$). The latter scenario is of particular interest as it is relevant to the experiments on MOES. The low energy properties of the system in this regime can be described by using the hydrodynamic theory of superfluids. Assuming a linear dependence of the local chemical potential on the 2D density, one recovers the results obtained in the Thomas-Fermi approximation, with the only difference being the value of the coupling constant $V_0$. Though one still has $\nu\ll 1$ for the power of the asymptotic in Eq. \eqref{onebodydensity}, the healing length $\xi$ itself is microscopically small, so that there is no extended coherence in a uniform configuration. We shall see, that the spontaneous onset of ODLRO in this regime becomes possible in the fragmented-condensate solid phase, where the originally 1D system breaks into a sequence of independent 2D harmonically trapped condensates.     

\subsection{One-dimensional mean field regime}
\begin{figure}[t]
\includegraphics[width=\columnwidth]{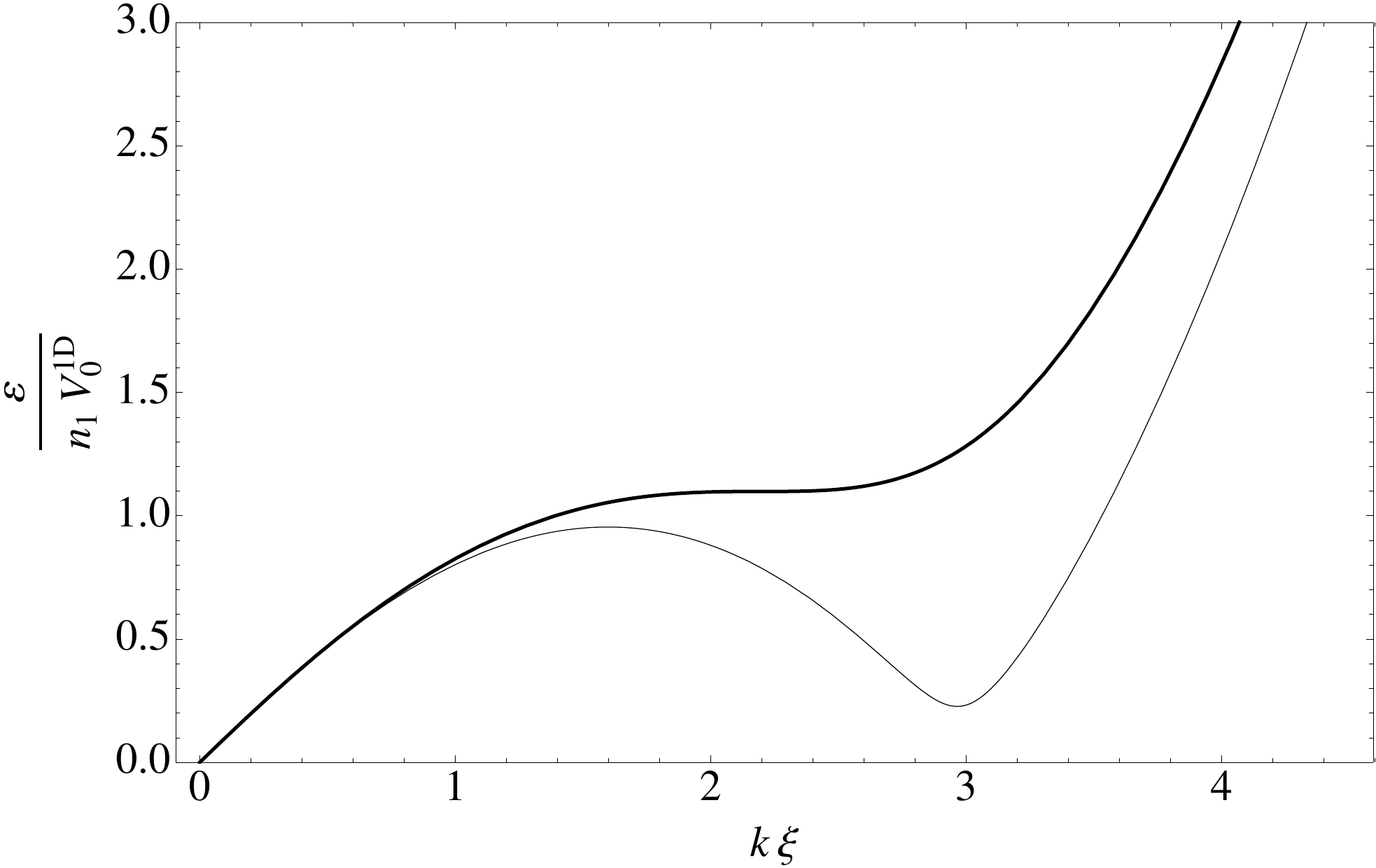}
\caption{Change of the shape of the elementary excitation spectrum of a dilute 1D dipolar condensate produced by the variation of the density $n_1$. We use Eq. \eqref{spectrum} with $\zeta=50$. For $n_1=n_{r}\equiv[2x_\ast\ln(\zeta/e^{3/2})]^{-1}$ the spectrum starts to develop a roton-maxon structure (solid thick line). An increase of $n_1$ pushes the roton minimum towards zero, as is illustrated by the thin line taken at $n_1=(n_r+30n_c)/31$.  At $n_1=n_c\equiv[2x_\ast \ln(\zeta/2e)]^{-1}$ the roton touches zero and for larger densities the uniform condensate becomes dynamically unstable.}
\label{Fig1}
\end{figure} 

In this regime the function $\varphi$ takes the form
\begin{equation}
\label{oscillator}
\varphi(y)=\pi^{-1/4}e^{-y^2/2a_y^2}.
\end{equation}
In the long-wavelength limit $k_x a_y\ll 1$ the first perturbative correction to the effective interaction \eqref{V01D} is provided by the anomalous part of the off-shell scattering amplitude\cite{Andreev3}
\begin{equation}
\label{1Dscattering}
f(k_x,k_x')=\frac{\hbar^2}{mx_{\ast}}(\lvert k_x - k_x'\rvert x_{\ast})^2\ln(\lvert k_x - k_x'\rvert x_{\ast}).
\end{equation}
By plugging \eqref{1Dscattering},  \eqref{oscillator} and \eqref{Contact} into the secondly-quantized Hamiltonian \eqref{SpinlessHamiltonian} one obtains
\begin{equation}
\label{Hamiltonian}
\begin{split}
\hat H_\ast=&\sum_{k_x}(E_{k_x}+\hbar\omega_y/2)\hat c^\dagger_{k_x}\hat c_{k_x}+\\
&+\frac{V_0^{1D}}{2L}\sum_{k_x, p_x, q_x}[1+\zeta(\lvert p_x-q_x\rvert x_\ast)^2\ln(\lvert p_x-q_x\rvert x_\ast)]\times\\
&\times\hat c^\dagger_{k_x+p_x}\hat c^\dagger_{k_x-p_x}\hat c_{k_x+q_x}\hat c_{k_x-q_x},
\end{split}
\end{equation}
where $E_k=\hbar^2 k^2/2m$ and we have introduced a dimensionless quantity
\begin{equation}
\zeta=\frac{\hbar^2}{mx_{\ast}V_0^{1D}}.
\end{equation}
The standard Bogoliubov approach then yields the elementary excitation spectrum in the form 
\begin{equation}
\label{spectrum}
\varepsilon(k)=\sqrt{E_k^2+2n_1V_0^{1D}E_k[1+\zeta (kx_\ast)^2\ln(kx_\ast)]}.
\end{equation}
A supplementary to \eqref{MF} condition for the validity of the mean-field result \eqref{spectrum} is $n_1 x_\ast\ll 1$. For small wavevectors the excitations are sound waves $\varepsilon(k)=\hbar c_1k$ with
\begin{equation}
\label{c1} 
c_1=\sqrt{n_1 V_0^{1D}/m\hbar^2}
\end{equation}
being the sound velocity. For the 1D density $n_1$ larger than $n_\mathrm{r}\equiv[2x_\ast\ln(\zeta/e^{3/2})]^{-1}$ the spectrum develops a roton-maxon structure. At the point 
\begin{equation}
\label{nc}
n_1=n_c\equiv[2x_\ast \ln(\zeta/2e)]^{-1}
\end{equation}
the roton minimum touches zero and the uniform condensate becomes dynamically unstable. One would expect that at this point the system exhibits a second order quantum phase transition\cite{Pitaevskii, Pomeau} to a stripe state
\begin{equation}
\label{stripe}
\psi(x)=\sqrt{n_1}(\cos{\theta}+\sqrt{2}\sin{\theta}\cos{k_r x})
\end{equation}
with $\theta\ll 1$ being an increasing function of the density (the corresponding equation can be obtained from \eqref{theta} by putting $ g_3^{1D}= 0$) and
\begin{equation}
\label{kr}
k_r=x_\ast^{-1}e^{-1/4n_1 x_\ast-1/2}
\end{equation}
corresponding to the position of the roton minimum. 
However, in the frame of the model \eqref{SpinlessHamiltonian} the inverse compressibility $\kappa^{-1}=\partial\mu/\partial n_1$ of the state \eqref{stripe} is negative, which implies a collapse.

\subsection{Stable dilute quasi-1D supersolid}
\begin{figure}[t]
\includegraphics[width=\columnwidth]{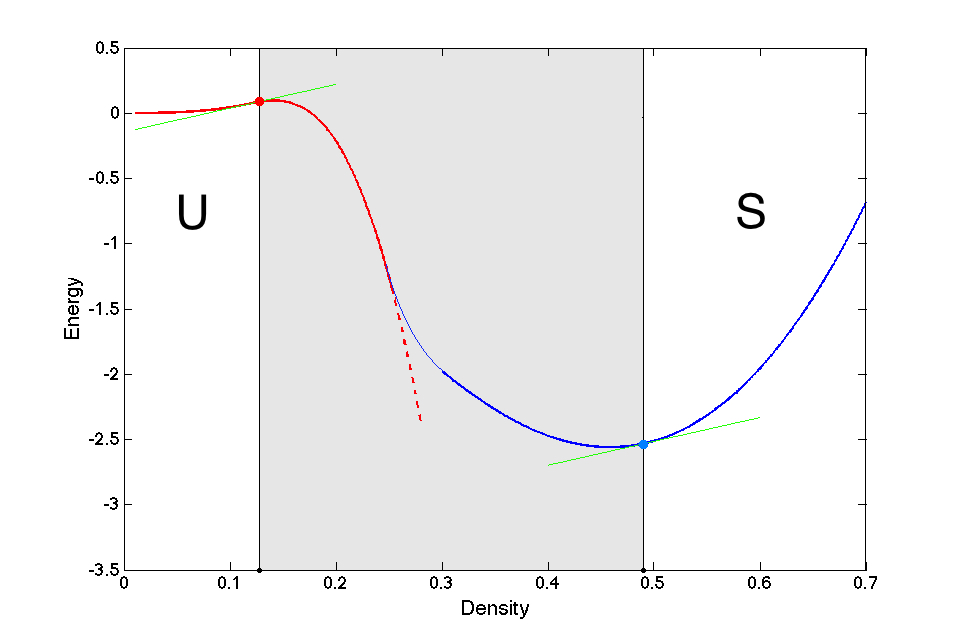}
\caption{The dimensionless energy $\tilde{\mathcal E}=\mathcal E x_\ast^2/V_0^{1D}$ of a quasi-one-dimensional condensate of bosonic dipoles as a function of the density $\tilde n_1=n_1x_\ast$. The uniform state (U) ceases to be the ground state at $ n_1= n_c^{(1)}$. At this point the system undergoes a first order quantum phase transition to a supersolid state (S) via a collapse (shaded area) accompanied by an increase of the instability magnitude $\theta$ and wavevector $k_0$. For small $\theta$ the corresponding dependence $\tilde{\mathcal E}(\tilde n_1)$ is given by \eqref{energy} with the substitution \eqref{theta} and is shown by the red line. As $k_0$ approaches $a_y^{-1}$, three-body repulsive forces become more efficient and, eventually, stabilize the system in a supersolid phase (S). The dependence of the energy of this phase on the density is given by \eqref{quasi1Denergy} and is shown by the blue line. The new stable point $n_c^{(2)}$ is defined by the requirement $\mu (n_c^{(2)})=\mu(n_c^{(1)})$, where $\mu=\partial\mathcal E/\partial n_1$ is the chemical potential of the system. We take $\zeta=10^4$, $x_\ast/a_y=0.1$ and $\gamma=1/4$ (for $\alpha=8$). We find $\tilde n_c^{(1)}=0.12$, $\tilde n_c^{(2)}=0.49$ and $k_0(n_c^{(2)})=0.48 a_y^{-1}$.}
\label{Fig2}
\end{figure}  

The collapse of the model \eqref{SpinlessHamiltonian} has already been observed for dipolar supersolids in the planar 2D geometry \cite{Collapse}. It has been shown that it is possible to stabilize the system by introducing three-body repulsive forces \cite{DiluteSupersolid}. One supplements the two-body Hamiltonian \eqref{SpinlessHamiltonian} with the term
\begin{equation}
\label{ThreeBody}
\frac{g_3}{\alpha}\int\hat\Psi(\bm\rho)^\dagger\hat\Psi(\bm\rho)^\dagger\hat\Psi(\bm\rho)^\dagger\hat\Psi(\bm\rho)\hat\Psi(\rho)\hat\Psi(\bm\rho)d\bm\rho,
\end{equation}
where the combinatorial factor $\alpha$ is equal to $3!=6$ for a spinless model. For the BCS-like model\cite{Andreev4} (see also Sec. IV) one should take $\alpha=8$, which corresponds to 6 possible ways of composing an interaction of an exciton and an excitonic pair (quasi-biexciton) in the system of equally populated spin-up and spin-down branches (a binary mixture).

Following the approach developed for 2D dipoles \cite{DiluteSupersolid}, we substitute the \textit{ansatz} \eqref{stripe} into the many-body Hamiltonian \eqref{SpinlessHamiltonian} supplemented with the term \eqref{ThreeBody} to obtain the energy density of the system
\begin{equation}
\label{energy}
\begin{split}
\mathcal E(k,\theta)=&\left(E_k n_1+\frac{\hbar^2n_1^2}{m x_\ast}(k x_\ast)^2 \mathcal D(k,\theta)\right)\sin^2\theta\\
&+V_0^{1D}n_1^2\mathcal C(\theta)+g_3^{1D}n_1^3\mathcal T(\theta),
\end{split}
\end{equation}         
where
\begin{subequations}
\begin{align}
\mathcal C(\theta)&=\frac{1}{32}(27-4\cos 2\theta-7\cos 4\theta)\\
\mathcal D(k,\theta)&=2\cos^2\theta \ln (k x_\ast)+\sin^2\theta \ln (2k x_\ast)\\
\mathcal T(\theta)&=\frac{1}{16\alpha}(55-15\cos 2\theta-27\cos 4\theta+3\cos 6\theta)
\end{align}
\end{subequations}
and $g_3^{1D}=g_3/\sqrt{3}\pi a_y^2$. Minimization of $\mathcal E(k,\theta)$ with respect to $k$ yields
\begin{equation}
\label{k01D}
\ln (k_0 x_\ast)=-\frac{2(n_1 x_\ast)^{-1}+3+2(1-\cos 2\theta)\ln 2+\cos 2\theta}{2(3+\cos 2\theta)}.
\end{equation}
For $\theta=0$ this gives the above result \eqref{kr}, and for $\theta=\pi/2$ one finds $\ln(2k_0 x_\ast)=-1/2n_1 x_\ast-1/2$ in agreement with the expression for the wavevector of a crystalline fluctuation of the exciton order parameter close to the critical temperature \cite{Andreev3}.

Substituting \eqref{k01D} into the energy functional \eqref{energy}, expanding the result up to the fourth power of $\theta$ and further minimizing it with respect to $\theta$ one obtains
\begin{equation}
\label{theta}
\theta^2=\frac{\zeta e^{-1/2n_1 x_\ast-1}-2-\frac{12g_3^{1D}}{\alpha V_0^{1D}} n_1}{(\frac{5}{3}+\ln 4+\frac{1}{2n_1 x_\ast})\zeta e^{-1/2n_1 x_\ast-1}-\frac{29}{6}-\frac{17g_3^{1D}}{\alpha V_0^{1D}} n_1},
\end{equation}
where the condition
\begin{equation}
\label{CriticalPoint}
\zeta e^{-1/2n_1 x_\ast-1}\geqslant 2+\frac{12g_3^{1D}}{\alpha V_0^{1D}} n_1 
\end{equation}
should be satisfied. The latter can be regarded either as a condition for the critical density $n_c^{(1)}$ [for $g_3^{1D}=0$ one recovers the result \eqref{nc}] or for the critical value of the contact interaction\cite{Andreev4} at which the transition to a supersolid state is expected to take place.

This transition is not, however, of the second (continuous) type, as one would expect on the general grounds\cite{Landau}. Indeed, as one can see in Fig. \ref{Fig2}, inclusion of the term \eqref{ThreeBody} does not save the situation for a purely 1D problem. At $n_1=n_c^{(1)}$ the crystalline order parameter $\theta$ spontaneously increases and the system transforms into a collapsing density wave (the corresponding dependence of $\mathcal E$ on the density for the typical values of the parameters is shown by the red line in the shaded area). The period of the wave decreases (the formula \eqref{k01D} with $\theta=\pi/2$) until it becomes on the order of the transverse oscillator length $a_y$.

As $k_0$ approaches $a_y^{-1}$ the two-dimensional origin of the system becomes important. The result \eqref{1Dscattering} for a purely 1D scattering process does no longer provide a good description of the interaction tail. Thus, for $k a_y\gg 1$ the anomalous correction to the scattering amplitude would behave as\cite{2Ddipoles}
\begin{equation}
\label{2Dscattering} 
f(\bm k,\bm k')\sim -\frac{\hbar^2}{m}\lvert\bm k-\bm k'\rvert x_\ast,
\end{equation}
where $\bm k$ is now a two-dimensional vector.

To describe the dipolar interaction in the condensate density wave in the crossover region $k_0 a_y\lesssim 1$ we shall employ the following expression
\begin{equation}
\label{quasi1Dscattering}
f(k_x,k_x')=\frac{2\sqrt{\pi}\hbar^2}{mx_{\ast}}\left(\frac{x_\ast}{a_y}\right)^2 U[-1/2,0,(\lvert k_x - k_x'\rvert a_y)^2/2],
\end{equation} 
where $U(a,b,x)$ is the Tricomi function. This expression can be obtained by properly averaging the 2D result \eqref{2Dscattering} over the  external confinement in the $y$ direction. Using \eqref{quasi1Dscattering} instead of \eqref{1Dscattering} and putting $\theta=\pi/2$ we obtain   
\begin{equation}
\label{quasi1Denergy}
\begin{split}
\mathcal E(k_0)=&E_{k_0} n_1+\frac{3}{2}V_0^{1D}n_1^2+\frac{5}{2\alpha}g_3^{1D}n_1^3\\
&-\frac{\sqrt{\pi}\hbar^2n_1^2}{2m x_\ast}\left(\frac{x_\ast}{a_y}\right)^2 U[-1/2,0,2(k_0 a_y)^2],
\end{split}
\end{equation}
where $k_0$ is the solution of the transcendental equation
\begin{equation}
\label{k0quasi1D}
e^{k_0^2 a_y^2} K_0(k_0^2 a_y^2)=(n_1 x_\ast)^{-1}
\end{equation}
with $K_n(x)$ being the modified Bessel function of the second kind. 

For large $k_0$ the equation \eqref{k0quasi1D} yields
\begin{equation}
\label{k02D}
k_0=\sqrt{\frac{\pi}{2}}\frac{x_\ast}{a_y}n_1,
\end{equation}
consistently with the analogous result for the 2D system \cite{DiluteSupersolid,Thesis}. In the same limit the equation for the energy density \eqref{quasi1Denergy} takes the form
\begin{equation}
\mathcal E=\frac{3}{4}V_0^{1D}n_1^2+g_3^{1D}n_1^3\left(\frac{5}{2\alpha}-\gamma\right),
\end{equation} 
where
\begin{equation}
\gamma=\frac{\sqrt{3}\pi^2\alpha}{64}\frac{\hbar^2 x_\ast^2}{mg_3}.
\end{equation}
One can see, that for sufficiently small $\gamma$ the compressibility $\kappa$ of the density wave becomes positive, which means stabilization of the system. The lower bound for $\gamma$ is defined by the requirement for the supersolid energy \eqref{quasi1Denergy} to be lower than the energy of the uniform state (U). Thus, we obtain
\begin{equation}
\label{gamma}
\frac{3}{2\alpha}<\gamma<\frac{5}{2\alpha}
\end{equation}
for the range of the parameter $\gamma$ where a stable supersolid state (S) may exist. 

The characteristic dependence of the energy of the quasi-1D supersolid on the density is shown in Fig. \ref{Fig2} by the blue line. We take $\alpha=8$ and $\gamma=1/4$. The shaded area corresponds to the collapse. In practice, one may expect that at the critical point $n_c^{(1)}$ defined by \eqref{CriticalPoint} (red point) there is a jump in the density of the system from $n_c^{(1)}$ to some value $n_c^{(2)}$ (blue point), which corresponds to the same chemical potential $\mu=\partial \mathcal E/\partial n_1$ (the tilt of the green tangents), but now in the stable supersolid phase. In other words, the U-S phase transition is of the first order.     
     
We now discuss possible realization of our model in physical systems. For ultra-cold gases of non-reactive NaK molecules the Hamiltonian \eqref{SpinlessHamiltonian} with the three-body term \eqref{ThreeBody} can be realized in the bilayer geometry with interlayer tunneling\cite{Petrov}. The effective three-body repulsion appears as a result of interaction of the third particle with a virtually excited interlayer bound state when the tunneling amplitude $t$ approaches some critical value $t_c$. In the same limit, the two-body interaction $V_0$ becomes proportional to $t-t_c$. In order to achieve the mean-field quasi-1D configuration for reasonable values of $x_\ast$, we propose to use a superposition of a standard 2D optical lattice and a 1D photonic crystal-based subwavelength lattice\cite{Subwavelength}. The 2D optical lattice generates an array of 1D tubes having an elliptic cross-section with the largest radius $a_y\sim 100$ nm ($\sim$10 kHz). The 1D photonic crystal with the period $d\sim10$ nm is then placed in the immediate vicinity of one tube in such a way as to split the tube into two stripes. The in-plane width of each stripe is $2a_y$ and their transverse width is on the order of $d$. The critical tunneling between the stripes should reach 1 $\mu$K, which certainly would be much larger than the chemical potential $\mu$ - the latter must not exceed $\hbar\omega_y/2\sim100$ nK. For $\alpha=6$ the stability region \eqref{gamma} corresponds to $g_3\sim 9\sqrt{3}\pi^2\hbar^2 x_\ast^2/32m$. For vanishing $V_0$ this quantity defines the critical density $n_c$ of the first order transition according to \eqref{CriticalPoint}. Taking $x_\ast\sim d$ one finds $n_c x_\ast^2\sim 0.01$, which yields $\mu\sim 1$ nK. In the stable supersolid phase the density would be about a factor of 5 larger. Then at the temperature $T\sim 1$ nK we would be well below the corresponding Berezinskii-Kosterlitz-Thouless (BKT) point, and at such temperature the system would form a quasi-regular 1D array of beads, each containing about 60 molecules (in each layer). By using the relationship \eqref{TDmu} for the 1D sound velocity and Eq. \eqref{nu}, we obtain $\nu\sim 3\sqrt{\pi} x_\ast/16 a_y$ for the exponent of the characteristic power law \eqref{onebodydensity} for the one-body density matrix. For our configuration we find $\nu\sim 0.03$, so that the coherence is extended over multiple periods of the structure.

Quasi-1D gases of dipolar excitons can be realized at the interfaces between the electron and hole regions in GaAs quantum wells\cite{Formation,Andreev5} (see Section III). The two-body interaction between the excitons can be reduced by using a resonance in the channel where the two excitons have opposite spins\cite{Andreev4,BCSmodel}. The effective three-body repulsive force may appear due to interaction of the third exciton with a weakly-bound exciton pair (biexciton) in the same limit where $V_0$ becomes vanishingly small. However, in contrast to the molecular setting discussed above, nothing is presently known about the dependence of this force on the detuning and its order of magnitude. Analysis of this problem would constitute an interesting subject for future work.

For equal populations of the exciton spin branches the resonantly paired model is reduced to \eqref{SpinlessHamiltonian} and \eqref{ThreeBody} with $\alpha=8$. At the typical electron and hole densities one has\cite{Andreev5}  $\hbar\omega_y\sim 1$ $\mu$eV, which corresponds to $a_y\sim 600$ nm. For the exciton dipolar length $x_\ast\sim50$ nm we are in the same range of the dimensionless parameters as for polar molecules, with $nx_\ast^2\sim 0.01$ being a good choice for the 2D density. In the stability region \eqref{gamma} and at $V_0=0$ the chemical potential $\mu\sim 0.1$ $\mu$eV. The cells of the supersolid are $\sim 2$ $\mu$m in the longitudinal direction and contain about 90 particles. With the existing experimental facilities\cite{High2012} it would be possible, in principle, to resolve such structure. However, cooling of the system down to $0.1$ $\mu$eV (in order to achieve large coherence length in quasi-1D geometry) is challenging. The lowest bath temperature achieved in the current experiments\cite{High2012} is around 0.1 K.
     
\subsection{2D cigar}

The transverse density profile of the cigar has the typical form of an inverted parabola
\begin{equation}
\label{parabola}
\varphi^2 (y)/a_y=(\mu-m\omega_y^2 y^2/2)/V_0n_1,
\end{equation}  
where $\mu$ is given by \eqref{muTF}. We omit the standard unit step function cut-off on the r.h.s. for brevity. As before, we expect the longitudinal density distribution to be somehow affected by the long-range dipole-dipole interaction. However, the perturbative method used in the dilute regime does not apply here. Instead, one can use phenomenological arguments.  

The result \eqref{1Dscattering} suggests that the dipolar tail can give rise to negative Fourier components $V^\mathrm{TF}(k)$ of the effective 1D interaction in some range of the momentum transfer. Hence, we take $V^\mathrm{TF}(k)$ in the form
\begin{equation}
\label{V}
V^\mathrm{TF}(k)=V_0^{TF}+V_\ast^\mathrm{TF}(k),
\end{equation}
where $V_\ast^\mathrm{TF}(0)=0$ and $V_\ast^\mathrm{TF}(k)<0$ for $k>0$. The contact part
\begin{equation}
V_0^{TF}=\frac{2}{5}\left(\frac{3}{2}\right)^{2/3}\left(\frac{mV_0}{\hbar^2}n_1 a_y\right)^{-1/3}\frac{V_0}{a_y}
\end{equation}
is obtained by substituting \eqref{parabola} into the Hamiltonian \eqref{SpinlessHamiltonian} and integrating over $y$. In the spirit of the seminal work on coherent crystals\cite{Kirzhnits} one can then suppose that $V^\mathrm{TF}(k)$ has a minimum at some $k=2k_0$ such that $k_0\ll \sqrt{n}$. We shall further assume $k_0^{-1}\gtrsim 2R_c$, which is consistent with the estimate of the wavelength of a dilute supersolid made in the previous subsection.

A natural choice for the ground state wavefunction would be
\begin{equation}
\label{cos}
\psi(x)=\sqrt{2n_1}\cos(k_0 x),
\end{equation}     
which formally corresponds to \eqref{stripe} at $\theta=\pi/2$. In terms of $\psi$ the energy of the system can be written in the form
\begin{equation}
\label{TFenergy}
\begin{split}
E_\mathrm{TF}[\psi]=&\frac{V_0^\mathrm{TF}}{2}\int \lvert\psi(x)\rvert^4dx+\frac{V_0^\mathrm{TF}n_1}{4}\int \lvert\psi(x)\rvert^2 dx\\
&+\frac{1}{2}\int V_\mathrm{auto}[\psi] \lvert\psi(x)\rvert^2dx,
\end{split}
\end{equation}
where we have introduced
\begin{equation}
\label{auto}
\begin{split}
V_\mathrm{auto}[\psi]&\equiv\frac{1}{2\pi}\int\int\ V_\ast^\mathrm{TF} (k) e^{ik(x-x')}\lvert\psi(x')\rvert^2dk dx'\\
&= n_1 V_\ast^\mathrm{TF}(2k_0) \cos(2k_0 x).
\end{split}
\end{equation}
The second term in \eqref{TFenergy} is due to the external trapping along $y$.

Eq. \eqref{TFenergy} may be regarded as the energy of a crystalline structure where the periodical lattice field \eqref{auto} and the periodical density distribution \eqref{cos} maintain each other in a self-consistent way. In the equilibrium the virial relation should hold between the contact interaction energy, external confinement and the energy due to autolocalization. By performing a scaling transformation
\begin{equation*}
\Psi'(x,y)=(1+\nu)\Psi[(1+\nu)x,(1+\nu)y]
\end{equation*} 
and imposing that the variation of the total energy \eqref{TFenergy} vanishes at first order in $\nu$, one gets the identity
\begin{equation}
\label{virial}
\frac{V_0^\mathrm{TF}}{2}\int \lvert\psi(x)\rvert^4dx=\frac{(V_0^\mathrm{TF}-2V_\ast^\mathrm{TF})n_1}{4}\int \lvert\psi(x)\rvert^2 dx
\end{equation}
Substitution of \eqref{cos} into \eqref{virial} and integration over the lattice period then yields
\begin{equation}
\label{virialV}
V_\ast^\mathrm{TF}(2k_0)=-V_0^\mathrm{TF},
\end{equation}
which sets the equilibrium values of energy \eqref{TFenergy} and chemical potential of the supersolid equal to those of the uniform configuration.

Providing that the condition \eqref{virialV} of mechanical equilibrium is satisfied, the function \eqref{cos} formally corresponds to a solution of the 1D Gross-Pitaevskii equation
\begin{equation}
\label{LDA}
V_0^\mathrm{TF}\psi^2(x)+V_0^\mathrm{TF}n_1/4+V_\mathrm{auto}(x)=\mu,
\end{equation}
obtained by minimization of \eqref{TFenergy} with respect to $\psi^\ast$. It is clear, however, that this picture becomes inadequate for the depleted regions, where $V_\mathrm{auto}(x)+V_0^\mathrm{TF}n_1/4$ approaches $\mu$. In these regions one should take into account the quantum kinetic energy which we have neglected so far. A straightforward way to do this we propose below is to approximate the lattice by a chain of coupled harmonic potentials and then proceed from the Thomas-Fermi solution of the problem \cite{Boundary}.

\subsection{Inhomogeneous state of the cigar as a chain of 2D trapped Bose-Einstein condensates}
\begin{figure}[t]
\includegraphics[width=1\columnwidth]{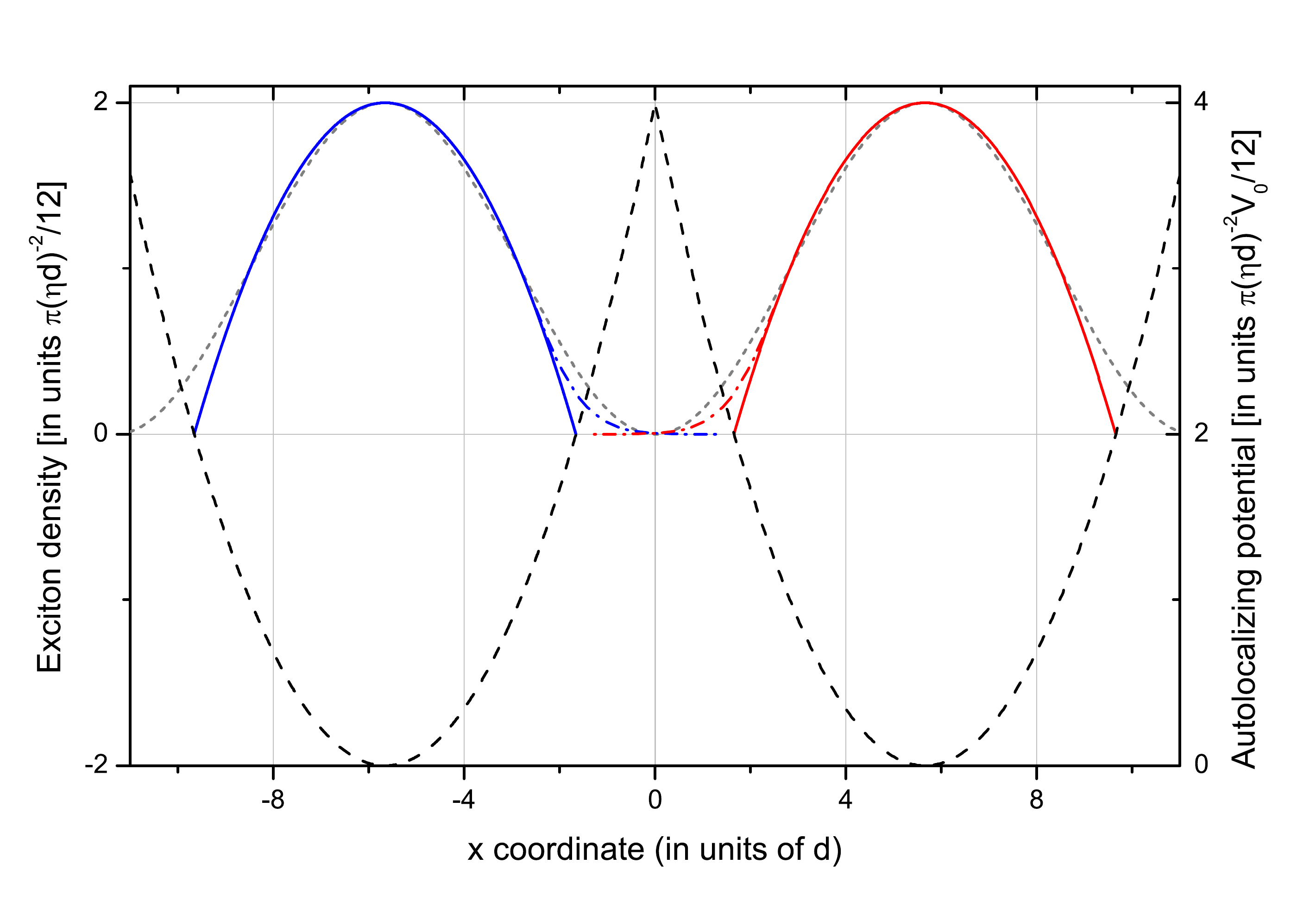}
\caption{Inhomogeneous state of the 2D cigar as a chain of trapped 2D Bose-Einstein condensates. The maxima of the density wave \eqref{cos} [short dash grey line] are approximated by the Thomas-Fermi paraboloids \eqref{2DLDA} (red and blue solid lines). The oscillator frequencies $\omega_j$ are chosen such that the corresponding harmonic traps intersect the cosine profile \eqref{auto} at the bending points defined as $\partial^2 V_\mathrm{auto}(x)/\partial x^2=0$. In the $y$-direction the width of the condensates is equal to the width of the cigar. In the boundary region in between the adjacent beads the Thomas-Fermi approximation fails. In this region the density profile takes universal form (blue and red dash-dot lines) which does not depend neither on the actual shape of the autolocalizing potential $V_{auto}(x,y)$, nor on the parameter $V_0$ characterizing the contact part of the exciton-exciton interaction \cite{Boundary}. These enter the expression for the characteristic length $d_x=(2m\vert F_x\vert/\hbar^2)^{-1/3}$ (which we conveniently use as a unit length), and the transformation $\Pi_j(x,y=0)=\sqrt{\pi/12}(\eta d_x)^{-1}\pi_{k_0}(x/d_x)$ for the trial function. We have taken $d_x=R_x/4$ in order to better visualize the boundary region in such specific scale. In practice, one expects $d_x\ll R_x$.}
\label{Fig3}
\end{figure}

Turning back to the two-dimensional representation, we replace the product of \eqref{cos} and \eqref{parabola} by
\begin{equation}
\label{Chain}
\Psi_{k}(x,y,t)=\sum\limits_{j=1}^{J}\Pi_j(x,y)e^{i\theta_j(t)},
\end{equation}
where $J=Lk/\pi$ is the total number of periods in the quantization length $L$ and the functions $\Pi_j(x,y)$ are given by
\begin{equation}
\label{2DLDA}
\vert\Pi_j(x,y)\vert^2=\frac{\mu_j-m\omega_y^2 y^2/2-m\omega_j^2 (x-x_j)^2/2}{\tilde V_0}
\end{equation}
with the renormalized coupling constant $\tilde V_0=V_0/2$ and satisfy the normalization condition
\begin{equation}
\int\vert\Pi_j(x,y)\vert^2 dxdy=N_j.
\end{equation}
Here $x_j=\sum\limits_{i=1}^{j}\lambda_i-\lambda_j/2$ are the positions of the density maxima along the chain, $\theta_j(t)$ are time-dependent phases of the local order parameters, $N_j$ are occupation numbers of different sites and $\mu_j$ are the corresponding chemical potentials. In the equilibrium $\theta_j(t)=\mu_j t$ and
\begin{equation}
\label{muj}
\begin{split}
&\mu_1=\mu_2=...=\mu,\\
&N_1=N_2=...=N/J\equiv N_0,
\end{split}
\end{equation}
where $\mu$ is given by \eqref{muTF} and $N_0$ will be specified below. The quantities $\lambda_i$ give the longitudinal sizes of the lattice sites. In the absence of disorder
\begin{equation}
\label{kj}
\lambda_1=\lambda_2=...= \lambda_0
\end{equation}
with $\lambda_0=\pi/k_0$ being the wavelength of the density wave in the original model (see above).    

The result \eqref{2DLDA} has been obtained by variation of the energy functional \eqref{TFenergy} over $\Psi^\ast\equiv\psi^\ast\phi^\ast/\sqrt{a_y}$ and has a simple form of the Thomas-Fermi paraboloid. Renormalization of the coupling constant physically reflects redistribution of the dipolar interaction energy in the density wave: half of this energy transforms into the energy due to self-trapping in the longitudinal direction.

In order to model the sites of the self-trapping potential $V_\mathrm{auto}(x)$ we take parabolas which intersect the cosine profile \eqref{auto} at the bending points defined as $\partial^2 V_\mathrm{auto}(x)/\partial x^2=0$. This eventually proves to be a better approximation than just Taylor expanding $\cos(x)$ function up to a quadratic term near its minima. We obtain
\begin{equation}
\omega_j^2=16 \mu_j/\lambda_j^2 m
\end{equation}
with $\mu_j$ and $\lambda_j$ given by Eqs. \eqref{muj} and \eqref{kj}, respectively. By definition of $\lambda_j$ one has
\begin{equation}
\label{continuity}
\sum\limits_{j=1}^{J}\sqrt{\frac{16\mu}{m\omega_j^2}}=L,
\end{equation}
the sum on the l.h.s being just a sum of the sizes of the cells.

The requirement for the chemical potential of the chain to be equal to that of the original model (i. e., to that of the cigar) yields the average number of particles per unit cell $N_0\approx 0.83 n_1\lambda_0$. Also the energy appears to be reduced by the factor $1.1$ with respect to the energy of the cigar. These minor deficiencies of our approximate solution are, however, fully recompensed by the usefulness of the Thomas-Fermi model \eqref{2DLDA} we arrived at.               

Indeed, the \textit{ansatz} \eqref{Chain} is the generalization of the well known double-well potential problem to the case of multiple wells, which has been extensively studied in the context of atomic BEC's in optical lattices \cite{Boundary, Lattices}. The role of the lattice potential here is played by an effective mean-field repulsive potential produced by the electrostatic interaction of a condensate with the corresponding neighbor. 

In the boundary region in between the adjacent condensates, where the Thomas-Fermi approximation fails, the density profile takes universal form which does not depend neither on the actual shape of the self-trapping potential, nor on the parameter $\tilde V_0$ characterizing the contact part of the interaction \cite{Boundary}. These enter the expression for the characteristic length $d_x=(2m\vert F_x\vert/\hbar^2)^{-1/3}$ (with $F_x=-\partial (m\omega_j^2 (x-x_j)^2/2)/\partial x$ being evaluated at the classical turning point defined by $m\omega_j^2 (x-x_j)^2/2=\mu$). We conveniently use $d_x$ as a unit of length in Fig. \ref{Fig2}, where we show two sites of the chain. Tiny overlap between the order parameters $\Pi_{j+1}$ and $\Pi_{j}$ locks the relative phases $\Phi_j=\theta_{j+1}-\theta_{j}$ and, at the absolute zero temperature, establishes the coherence over the distance $x\gg\ \lambda_0$. The stationary state of the chain corresponds to the symmetric configuration with $\Phi_j$ being equal to zero.

The dynamics of the chain can be described by $2(J-1)$ coupled equations of motion (Josephson equations) \cite{Josephson, Boundary},
\begin{equation}
\label{DWequation1}
\frac{\partial\Phi_j}{\partial t}=-\frac{E_C}{\hbar}k_j,
\end{equation}
for the relative phases and 
\begin{equation}
\label{DWequation2}
\hbar\frac{\partial k_j}{\partial t}=E_J\sin\Phi_j,
\end{equation}
for the relative numbers of particles $k_j=(N_{j+1}-N_j)/2$. Here
\begin{equation}
\label{DWamplitude}
E_J=\frac{\hbar^2}{m}\int dy\left(\Pi_{j+1}\frac{\partial\Pi_{j}}{\partial x}-\frac{\partial\Pi_{j+1}}{\partial x}\Pi_{j}\right)_{x=0}
\end{equation}
is the amplitude of the oscillating particle currents due to deviations of the local chemical potentials $\mu_j=\partial E/\partial N_j$ from their equilibrium value $\mu$ and
\begin{equation}
\label{Ec}
E_C=2\frac{d\mu_j}{dN_j}
\end{equation}          
is the relevant interaction parameter of the problem. Both quantities should be evaluated at $N_j=N_0$ and, therefore, do not depend on $j$. By using the Thomas-Fermi model \eqref{2DLDA} one can obtain
\begin{equation}
\label{EcEvaluated} 
E_C=\frac{\mu}{N_0}
\end{equation}
and
\begin{equation}
E_{J}\sim N_{0}^{-1/6}k_{B}T_{c}^0e^{-S},
\label{EJ}
\end{equation}
where $k_BT_c^0=\hbar(6/\pi^2 N_0\omega_x\omega_y)^{1/2}$ is the critical temperature of BEC in an ideal harmonically trapped gas and $S\sim(H-\mu)/\hbar(\omega_x\omega_y)^{1/2}$ with $H$ being the height of the barrier\cite{Rontani,Zapata}. The latter, as one can see in Fig. \eqref{Fig3}, is twice the chemical potential. Taking into account this important property of our model, one can estimate the ratio $E_C/E_J$ as
\begin{equation}
\label{CoherenceRegimes}
\frac{E_{C}}{E_{J}}\sim\frac{\eta\exp(\eta\sqrt{N_{0}})}{N_{0}^{5/6}},
\end{equation}
where we have introduced
\begin{equation}
\label{eta}
\eta=\frac{\mu}{k_B T_c^0}=\sqrt{\frac{\pi}{6}\frac{m \tilde V_0}{\hbar^2}}.
\end{equation}
The condition for applying the mean-field approach to describe the chain of Josephson junctions reads
\begin{equation}
\label{CoherentRegime}
\frac{E_{C}}{E_{J}}\ll1
\end{equation}
and corresponds to the coherent regime where the number of particles in each site $N_j$ exhibits strong fluctuations $\langle \Delta N_{j}^2\rangle\gg 1$ around the equilibrium value $\langle N_j\rangle\equiv N_{0}$. Provided that typically $N_{0}\sim 10^{3}$, one can see that the condition \eqref{CoherentRegime} is well satisfied for $\eta\leqslant 0.2$.

Finally, let us note that the equations \eqref{DWequation1} and \eqref{DWequation2} can be recast in the canonical form
\begin{equation}
\label{CanonicalForm}
\begin{split}
\frac{\partial\Phi_j}{\partial t}&=\frac{\partial H_J}{\partial(\hbar k_j)}\\
\frac{\partial(\hbar k_j)}{\partial t}&=-\frac{\partial H_J}{\partial\Phi_j},
\end{split}
\end{equation}
where
\begin{equation}
\label{JosephsonHamiltonian}
H_J=E_C\sum\limits_{j=1}^J\frac{k_j^2}{2}-E_J\sum\limits_{j=1}^J \cos\Phi_j
\end{equation}
is the Josephson Hamiltonian. This form allows one to identify $\Phi_j$ and $\hbar k_j$ as the natural canonically conjugated variabes of the problem.

\subsection{Beyond mean field. Fragmentation of the supersolid}

We shall now deepen our understanding of the dipolar supersolid by proceeding from the "chain" model introduced above. We have argued that this model provides adequate description of the coherent density wave. Provided that the condition \eqref{CoherentRegime} is satisfied, the excitons are delocalized, i. e., they share the same wave function \eqref{cos}. The coherence length is governed by the long-wave quantum fluctuations of the phase. Further insight can be gained by exploring what happens when $E_{C}\sim E_{J}$ and $E_C>E_J$. To study this regime one should go beyond the classical mean field description of the Josephson effect.

The appearence of the depleted regions in the condensate due to its dynamical instability brings about phase fluctuations of entirely different nature than those usually considered for uniform systems. These are the quantum fluctuations of the relative phases between the macroscopic domains, each of which can be assumed to be fully coherent at $T=0$. The fluctuations of this type can be introduced into the problem by quantizing the Josephson equations \eqref{CanonicalForm} describing the coupling between the domains. The quantization can be achieved in the usual way by replacing the conjugated variables $k_j=(N_{j+1}-N_j)/2$ and $\Phi_j=\theta_{j+1}-\theta_j$ with operators satisfying the commutation relation \cite{Fluctuations}
\begin{equation*}
[\hat\Phi_j,\hat k_j]=i.
\end{equation*}
As a result, the observables $\hat\Phi_j$ and $\hat k_j$ obey the uncertainty relation
\begin{equation}
\label{uncertainty}
\langle \hat k_j^2\rangle\langle\sin^2\hat\Phi_j\rangle\geqslant\frac{1}{4}\langle\cos\hat\Phi_j\rangle^2,
\end{equation}     
which is analogous to the uncertainty relation for the angle and angular momentum in the standard quantum mechanics \cite{Carruthers}. 

In the case of strong tunneling $E_J\gg E_C$ the values of the relative phases $\Phi_j$ are localized around $0$ and the uncertainty relation \eqref{uncertainty} takes the usual form
\begin{equation*}
\langle \hat k_j^2\rangle\langle\hat\Phi_j^2\rangle\geqslant\frac{1}{4}.
\end{equation*}
One can see that the coherence assumed in our mean field models manifests iteself in the strong quantum fluctuations of the number of particles in each site. The exciton order parameter is spread out over multiple periods of the chain.

As one can see from Eq. \eqref{EJ}, the tunneling amplitude $E_J$ can be tuned by changing the parameter $\eta$ which characterizes the contact part of the exciton-exciton interaction. An increase of the parameter $\eta$ leads to a decrease of the size of the overlapping region and to a growth of the self-trapping potential barrier. The adjacent condensates extrude each other, approaching the ideal Thomas-Fermi density profiles (solid lines in Fig. \ref{Fig2}).

It is known from the physics of granular superconductors \cite{Bradley}, that for $T=0$ the Josephson Hamiltonian \eqref{JosephsonHamiltonian} exhibits a quantum phase transition to a charge-ordered insulating state at $E_C\geqslant 1.62E_J$. By using the relation \eqref{CoherenceRegimes} and taking $N_0\sim 10^3$ one finds
\begin{equation}
\label{etac} 
\eta_c\approx 0.27 
\end{equation}
for the critical value of the interaction parameter $\eta$ defined in \eqref{eta}.

\subsection{Topological transformation of the fragmented-condensate solid conserving simultaneously the total number of particles, the interaction energy and the chemical potential}

\begin{figure}[t]
\includegraphics[width=1\columnwidth]{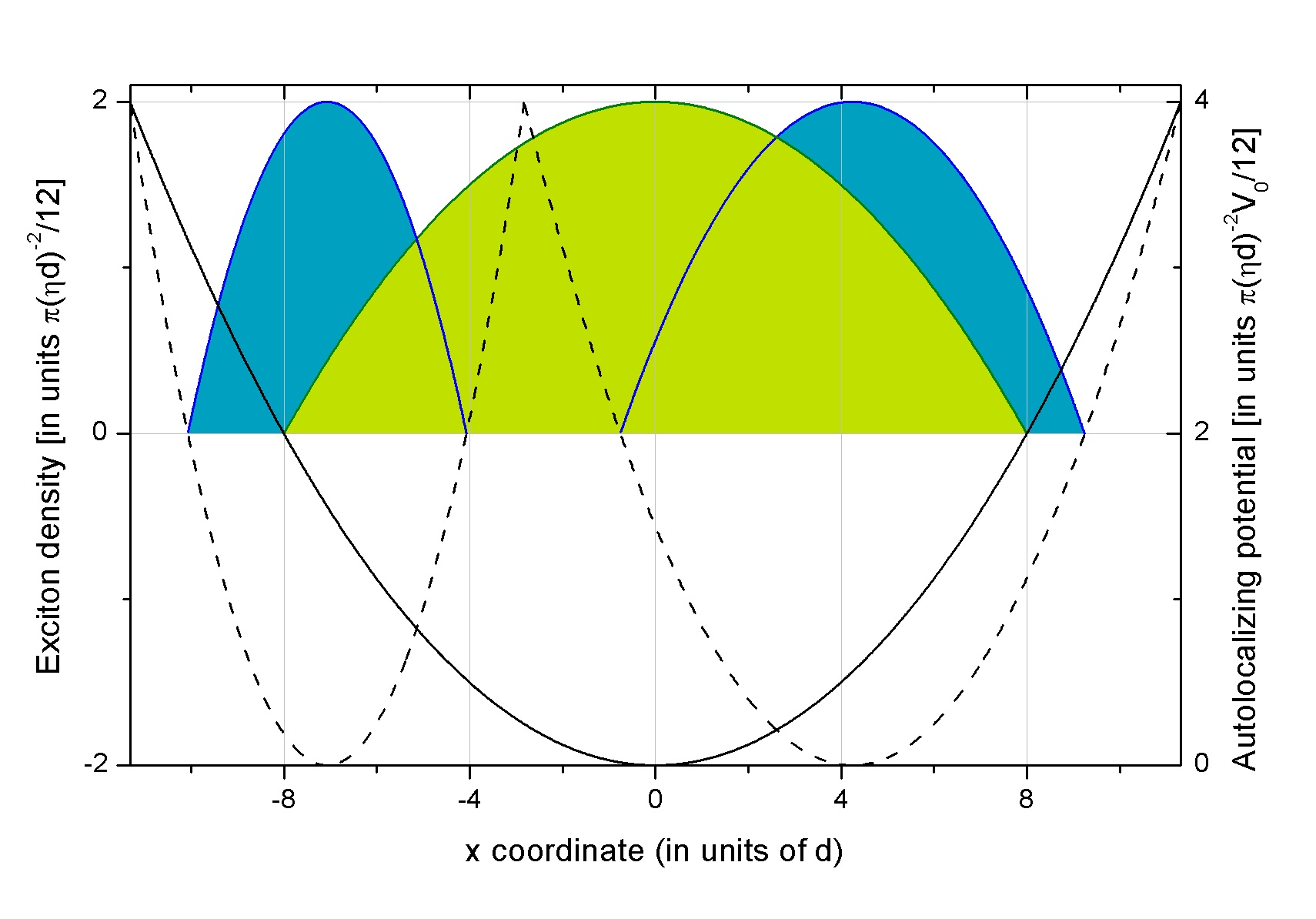}
\caption{Topological transformation of the fragmented-condensate solid conserving simultaneously the total number of particles $N$, the chemical potential $\mu$ and the total energy of the system $E$. In the Thomas-Fermi limit the density profile of each bead $\lvert\Pi_j(x,y)\rvert^2$ takes the form on an inverted paraboloid. The height of the paraboloid is defined by the chemical potential according to $\lvert\Pi_j(0,0)\rvert^2=\mu/\tilde V_0$. Providing that the condition \eqref{continuity} is satisfied, the total energy and the number of particles in the green paraboloid is the same as in a system of the two blue ones. We have used the same units of length, density and energy as in Fig. \ref{Fig3}.}
\label{Fig4}
\end{figure}           

A peculiar property of the chain model \eqref{Chain} is that the corresponding Thomas-Fermi energy \eqref{TFenergy} is invariant under an arbitrary change of the values of the wavelengths \eqref{kj} at the fixed chemical potential $\mu$.  Namely, one can show that the replacement
\begin{equation}
\{\omega_{j}\}\rightarrow \{\omega_{i}\}^{\ast },  
\label{transform}
\end{equation}
conserves simultaneously the total energy $E_\mathrm{TF}$ and the number of particles $N$. Here $\{\omega_{i}\}^{\ast}$ is a new set of oscillator frequencies, $i=1,2...I$, satisfying the continuity condition \eqref{continuity} with $i$ instead of $j$ and $I\neq J$. An example of such transformation is schematically shown in Fig. \ref{Fig4}.

A rigorous mathematical proof of this essentially topological result is given in Appendix A. The following remarks are in order here:
\begin{itemize}
\item In the absence of disorder, this property is inherited by the system from the original model \eqref{cos}.
\item The existence of an effective interaction of the type \eqref{V} is not crucial. In fact, by evaluating the mean-field interaction energy just of a chain consisting of $J$ inverted paraboloids, one can show that this energy is invariant under an arbitrary variation of $J$ and longitudinal sizes of the paraboloids, providing that one fixes the effective chemical potential (the height of the paraboloids) and the length of the chain\cite{Andreev1,Andreev2}. This can be especially relevant for driven-dissipative systems, where a seed periodical structure can be formed by an initial quench of $V_0$ and then grown by shifting a dynamical equilibrium with the reservoir.
\item Instead of a paraboloid, one can take any function of the type
\begin{equation}
\label{figure}
\frac{z}{h}=\left(\frac{x}{R_{x}}\right)^{2p}+\left(\frac{y}{R_{y}}\right)^{2p}
\end{equation}
for the condensate density profile. Here $z\leqslant h$ and $p$ is a non-zero natural number. This provides a possibility of a generalization of the theory accounting for the beyond mean-field corrections in the equation of state.  
\item At zero temperature the presence of such degeneracy implies that the lattice will tend to increase its average period $\lambda$ (reduce the number of sites) in order to minimize the quantum kinetic energy due to boundaries. The equilibrium value of $\lambda$ will be defined by the balance between the kinetic energy and an eventual change in the interaction energy $E_\mathrm{TF}$ due to violation of the condition \eqref{virialV}.
\item For $\lambda$ sufficiently close to $\lambda_0$ one can write  
\begin{equation}
E_\mathrm{TF}(k)=E_\mathrm{TF}(k_0)+\frac{dE_\mathrm{TF}(k_0)}{dk}(k-k_0)^2,
\end{equation}
where $k=\pi/\lambda$ and $k_0=\pi/\lambda_0$ corresponds to the minimum of the effective interaction \eqref{V}. Analogous expansion for the kinetic energy begins with the first power of $(k-k_0)$. Thus, there is a useful range of $\lambda$, where the change of the total energy of the supersolid is only due to variation of the kinetic energy.
\end{itemize}      

\subsection{Fragmented-condensate solid at a finite temperature (we assume $T_c^0/N\ll T\ll T_c^0$)}

The difference between the coherent and Fock regimes of the 1D chain becomes especially pronounced at a finite temperature. Here, the long-wave fluctuations of the phase would destroy the conventional supersolid, whereas the coherence length for the number-squeezed configuration would remain as large as the average size of a bead $\lambda$. Another important distinction is the dependence of $\lambda$ on the density $n_1$. In contrast to the result \eqref{k02D} characteristic for a dilute system, the wavelength of the fragmented-condensate solid turns out to be an \textit{increasing} function of $n_1$.  

At $T>0$ the appropriate thermodynamic potential is the free energy $F$. Assuming $\lambda$ being on the order of $\lambda_0$ we neglect possible variation of the interaction energy (see the previous subsection) and write the function $F$ in the form
\begin{equation}
F=E_\mathrm{kin}+F_\Phi,
\end{equation}
where
\begin{equation}
\label{kin}
E_\mathrm{kin}=\frac{\hbar^2}{2m}\int \lvert\nabla\Psi (\bm\rho)\rvert^2d\bm\rho
\end{equation}
is the kinetic energy (quantum pressure) and $F_\Phi$ is the part of the free energy due to the phase fluctuations. In the Thomas-Fermi approximation one can write (see Appendix B).
\begin{equation}
\label{QP}
E_\mathrm{kin}=J\frac{x}{\eta}k_B T_c^{0},
\end{equation}
where $J$ is the number of beads in the quantization length $L$ (it is related to the fragmented-condensate wavelength by $J=L/\lambda$) and $x$ is some dimensionless coefficient defined by the topology of the fragmented-condensate density profile in the depleted regions in between the beads. We assume $x$ to be not dependent on $J$.

The fluctuation part of the free energy $F_\Phi$ can be calculated starting from the Hamiltonian \eqref{JosephsonHamiltonian}. In the Fock regime this Hamiltonian reduces to
\begin{equation}
\label{reduced}
\hat H_J=-\frac{1}{2}E_C\sum\limits_{j=1}^J\frac{\partial^2}{\partial\Phi_j^2}.
\end{equation}
The corresponding eigenstates are plain waves
\begin{equation}
\label{planewave}
\Theta_{\{k_j\}}\sim \exp\left(i\sum\limits_{j=1}^J k_j\Phi_j\right)
\end{equation}
for a set of integer values $\{k_j\}$, so that the ground state function is a constant, revealing that the relative phases between the beads are distributed in a random way. According to the uncertainty relation \eqref{uncertainty} the variances $<k_j^2>$ of the number of particles $N_j$ in each site are instead vanishingly small - the fragments have well-defined number of excitons.

As one can see from \eqref{planewave}, the dynamics of the phase corresponds to mechanical motion of a free particle in J-dimensional space. Inserting a Josephson junction is thus equivalent to adding a new degree of freedom to the system. The partition function can be factorized and it takes the form
\begin{equation}
Z_\Phi=\sum_{\{k_j\}}e^{-\beta E_{\{k_j\}}}=\left( \sum_k e^{-\beta E_k}\right)^J,
\end{equation}
where $E_k=E_C k^2/2$ according to \eqref{reduced} and $E_C$ is given by \eqref{EcEvaluated}. Assuming $\beta E_C\ll 1$ we obtain
\begin{equation}
F_\Phi=-k_BT\ln Z_\Phi=-\frac{J k_B T}{2}\ln\left(\frac{2\pi}{\eta}\frac{T}{T_c^0}\frac{N}{J}\right).
\end{equation} 
Collecting both the kinetic energy and the fluctuation terms we write
\begin{equation}
\label{FreeEnergy}
F=J\frac{x}{\eta}k_B T_c^{0}-\frac{J k_B T}{2}\ln\left(\frac{2\pi}{\eta}\frac{T}{T_c^0}\frac{N}{J}\right)
\end{equation}
for the free energy of the fragmented condensate. Considered as a function of $J$, it has a minimum at the point
\begin{equation}
\label{BeadsNumber}
J=\frac{2\pi}{\eta e}\frac{T}{T_c^0}N\exp\left(-\frac{2x}{\eta}\frac{T_c^0}{T}\right).
\end{equation}
Note, that, by virtue of the relation \eqref{eta}, the parameter $T_c^0$ does not depend on $J$. It can be conveniently expressed as $k_B T_c^0=nV_0/\eta$, where $n$ is the average $2D$ density of the system at $T=0$. By using the relation $n_1/n=R_y$, we can rewrite the result \eqref{BeadsNumber} in the useful form
\begin{equation}
\label{lambda}
\lambda(n,T)=\frac{\eta e}{2\pi}\frac{V_0}{R_y k_BT}\exp\left(\frac{24x}{\pi}\frac{\hbar^2n/m}{k_BT}\right).
\end{equation}
Note, that though the $1/R_y$ prefactor decreases with $n$, the exponential factor overweighs this contribution, so that $\lambda$ \textit{increases} with $n$. This is in stark contrast with the result \eqref{k02D} obtained in the dilute regime.

\section{Comparison with the experiment}

A convenient setting for experimental study of quasi-1D quantum gases of dipolar excitons is the ring-shaped interface between the electron- and the hole-rich regions in biased semiconductor quantum wells (CW's). The regions of opposite charge may form under cw photoexcitation either in a single\cite{Combescot, BellLabs} QW or two coupled\cite{Butov2002, Pittsburgh} QW structures. The bias voltage $V_g$ applied in the transverse direction injects electrons into the nearest to the positive lead QW layer. In the absence of the photoexcitation these electrons fill uniformly the layer plane. A focused laser is then used to excite electron-hole pairs in the surrounding barriers. Holes are captured efficiently by the QW layer close to the negative lead and form a lake of a positive charge around the excitation spot. The interface between the hole lake and the outer unperturbed electron sea is seen as a bright ring in the exciton photoluminescence (PL).\cite{Formation}

Being remote from the hot photoexcitation spot the ring represents a source of cold excitons. An in-plane electric field induced at the interface due to the macroscopic separation of charge confines the in-plane translational motion of excitons in the radial direction\cite{Andreev5}. At $T>T_c$ the excitons are free to move along the interface, in between the so-called localized bright spots (LBS's) - point-like defects which pin the position of the ring\cite{Commensurability}. Thus, the PL ring consists of independent segments, having different lengths and widths due to the structural disorder. The segments are characterized by strong repulsive interaction due to built-in dipole moments of the excitons. The two-body repulsion ensures fast thermalization of excitons and makes possible observation of a Bose-Einstein condensed metastable state\cite{Thermalization}.

Below $T_c$ (few degrees K) all the segments simultaneously fragment into arrays of regularly spaced beads. This low-temperature state of the ring has been dubbed the Macroscopically Ordered Exciton State\cite{Butov2002} (MOES, see Introduction). In our previous Letter\cite{Andreev1} we have demonstrated consistency of the formula \eqref{lambda} for the MOES wavelength in the regime where the ring is allowed to expand when increasing the laser excitation power $P$ (the gate voltage $V_g$ is fixed). In that kind of experiment both the average exciton density $n$ and the width of the ring $2R_y$ remain roughly constant with $P$. The formula \eqref{lambda} then predicts a constant $\lambda$, which is what indeed has been observed.

Here we compare our results with another regime, in which the radius of the ring is kept fixed by simultaneously adjusting $V_g$ and $P$\cite{Commensurability}. In this regime both the density and the width of the MOES increase with $V_g$ (and $P$). The formula \eqref{lambda} correctly describes the behavior of $\lambda$ in these conditions as well. As a complementary test, we propose to measure the critical temperature $T_c$ (the temperature at which the MOES disappears) as a function of $V_g$.

\begin{figure}[t]
\includegraphics[width=1\columnwidth]{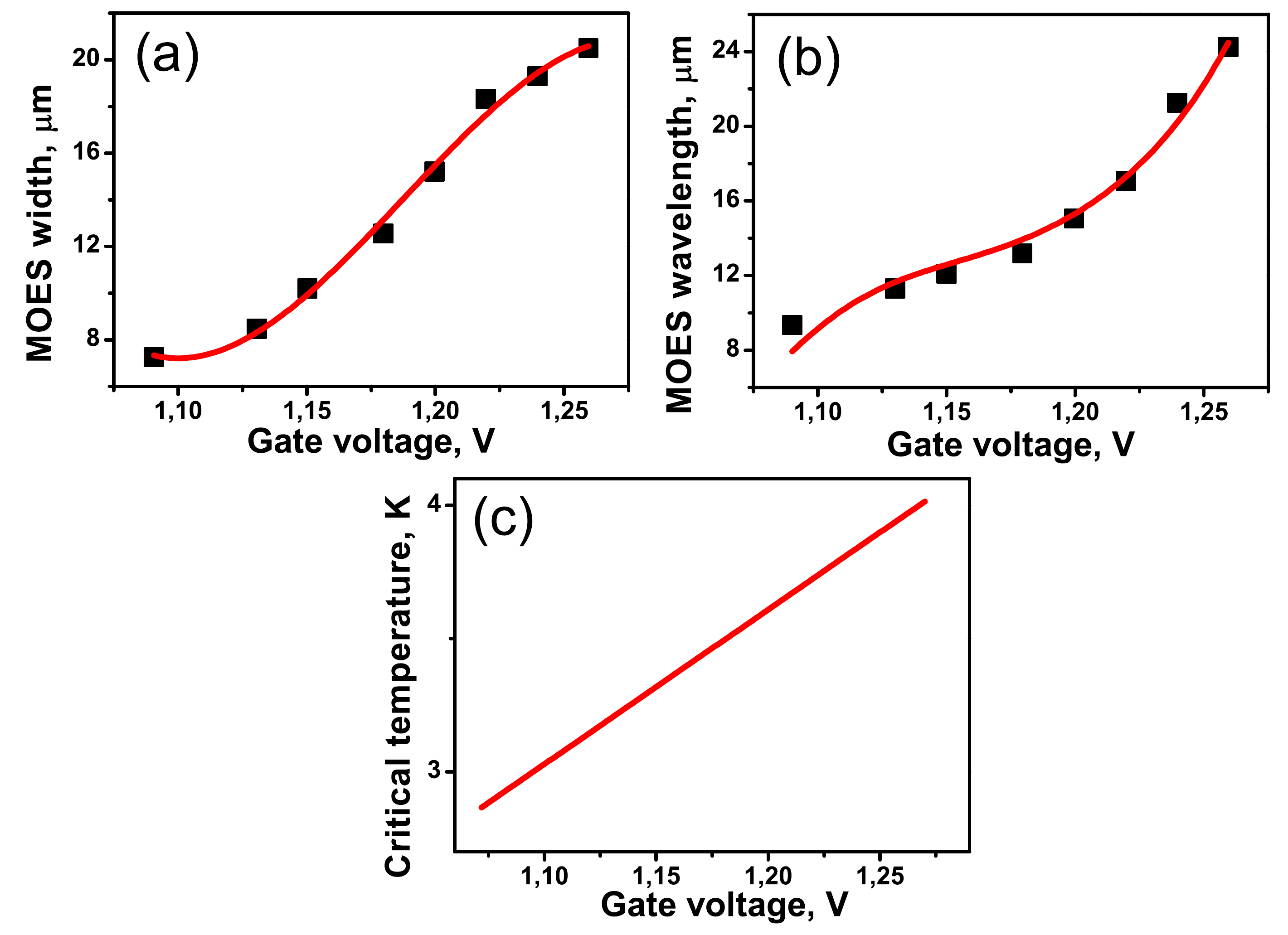}
\caption{Dependence of the width (a), the wavelength (b) and the critical temperature (c) of the MOES on the gate voltage $V_g$. Squares are the experimental data\cite{Commensurability}. In (a) the solid line is the cubic function \eqref{widthfit} with $c_3$, $c_2$, $c_1$ and $c_0$ being the fitting parameters. We substitute thus obtained analytic equation for the MOES width into Eq. \eqref{lambda}. The formula \eqref{lambda} is then used to fit the corresponding dependence of the wavelength in (b). We fix $x=4$ and $V_0=6.8$ $\mu eV\times\mu m^2$, consistently with the earlier estimates\cite{Andreev1,zimmermann}. The bath temperature is $T=1.6$ K. The only adjustable parameters are $n(V_g^\ast)$ and $dn(V_g^\ast)/dV_g$, which enter the expansion \eqref{densityfit} for the exciton density $n$ around $V_g^\ast=1.17$ V. The obtained dependence of $n$ on $V_g$ is used to plot the values of the critical temperature [Eq. \eqref{criticaltemp}] at which the MOES is expected to disappear [solid line in (c)].}
\label{Fig5}
\end{figure}     

\subsection{The width of the MOES}

In practice, the width of the MOES is determined by a complex interplay between the reaction-diffusion processes in the surrounding electron-hole plasma and the trapping of excitons at the ring. For our purposes it is sufficient to obtain a phenomenological fit of the experimentally measured dependence of $R_y$ on $V_g$, which we could then substitute into Eq. \eqref{lambda}, leaving a detailed discussion of the underlying physics for future work. A straightforward choice is a polynomial function. Good agreement with the experiment is achieved by using the cubic formula
\begin{equation}
\label{widthfit}
2R_y=c_3V_g^3+c_2V_g^2+c_1V_g+c_0.
\end{equation}
The result of the fitting procedure is shown in Fig. \eqref{Fig5} (a). Note a pronounced bend of the curve at low voltages. At the same voltages, there is also a bend in the corresponding plot of the MOES wavelength presented in Fig. \eqref{Fig5} (b). However, the direction of this bend is opposite to that observed for the MOES width.        

\subsection{The wavelength and the critical temperature of the MOES}

In order to compare the result \eqref{lambda} with the measured dependence of the MOES wavelength on $V_g$ we need to know the corresponding dependence for the exciton density $n$. Since the relative variation of $V_g$ in the experiment is small, we can write
\begin{equation}
\label{densityfit}
n(V_g)=n(V_g^\ast)+\frac{dn(V_g^\ast)}{dV_g}(V_g-V_g^\ast)
\end{equation}
where $V_g^\ast$ is some \textit{workpoint}. The choice of $V_g^\ast$ is arbitrary within the window used in the experiment and only slightly affects the values of the expansion coefficients $n(V_g^\ast)$ and $dn(V_g^\ast)/dV_g$ at which the best fit is achieved. We take $V_g^\ast=1.17$ V. Note, that the expansion of the type \eqref{densityfit} can hardly be justified for the dependence of $n$ on the excitation power $P$ - the latter varies over several orders of magnitude in the experiment\cite{Commensurability}. For this reason we do not consider the dependence of $\lambda$ on $P$ at all.

Substituting the phenomenological expression \eqref{widthfit} for $R_y$ with the coefficients $c_i$ determined in the previous subsection and the expansion \eqref{densityfit} for the exciton density into Eq. \eqref{lambda}, we fit the experimentally measured values of $\lambda$ presented in Fig. \eqref{Fig5} (b) by varying the values of $n(V_g^\ast)$ and $dn(V_g^\ast)/dV_g$. The bath temperature $T=1.6$ K is fixed. We also fix the parameters $x$ and $V_0$. We take $V_0=6.8$ $\mu eV\times\mu m^2$ which is $4$ times larger than the known estimate of this quantity \cite{zimmermann}. This yields $\eta=2.3$ according to Eq. \eqref{eta}. The reason for such choice of $V_0$ is that we expect tightly bound excitonic pairs (biexcitons) to form the supersolid media (see Section IV), with their dipole moments being twice the dipole moment of excitons. The dimensionless coefficient $x$ [see Eq. \eqref{QP} for the quantum pressure] should be on the order of unity. We take $x=4$ consistently with our previous estimates\cite{Andreev1}.

The obtained values of the parameters $n(V_g^\ast)$ and $dn(V_g^\ast)/dV_g$ can then be used to estimate the critical temperature $T_c$ at which the MOES disappears. By using the relation \eqref{eta} and assuming $\mu=n V_0$, one can recast the temperature of a non-interacting BEC in a 2D harmonic trap as\cite{footnote}
\begin{equation}
\label{criticaltemp}
k_B T_c^0=\frac{12\eta}{\pi}\frac{\hbar^2 n}{m},
\end{equation}
where $n$ depends on $V_g$ according to \eqref{densityfit}. By substituting the values of $n(V_g^\ast)$ and $dn(V_g^\ast)/dV_g$ found from the fitting of the MOES wavelength, we obtain the linear dependence of $T_c^0$ on $V_g$ shown in Fig. \ref{Fig5} (c). The obtained values of $T_c^0$ are on the order of the quantum degeneracy temperature of an exciton gas with $n\sim 10^{10}$ cm$^{-2}$.

\subsection{Commensurability of the MOES}

We close this section by an explanation of the recently observed commensurability of the MOES. As we have already mentioned (see Introduction), commensurability is inherent for supersolid states of matter. To see how this property is retained in our thermodynamic model of the fragmented-condensate solid, let us look at Eq. \eqref{FreeEnergy}. The stable configuration of the system is achieved when the minimum of the function \eqref{FreeEnergy} coincides with some integer value $J$ of Josephson junctions. If, by varying the control parameters, we shift the position of the minimum toward $J+1$, we shall certainly encounter the situation where the states with $J$ and $J+1$ junctions have the same free energy. In this case the system will exhibit strong fluctuations between these two states. In practice, this means that a well-defined pattern of beads can only be observed for some discrete values of the gate voltage.

\section{Discussion}

Our analysis of the effect of long-range dipolar repulsion on the Bose-Einstein condensation of a quasi-1D exciton gas can be sketched as follows. Consider what happens with the system, when one slowly decreases the temperature down to the temperature of quantum degeneracy. When approaching $T_c$ from above coherent fluctuations of the exciton order parameter $\Psi$ appear. In the fluctuation region the mean-field component of the system is very dilute, and the main conclusions drawn in Subsection A can be applied. The predicted roton instability here would reveal itself as a non-conventional fluctuation of $\Psi$ having a crystalline structure\cite{Andreev3}. Importantly, for such fluctuations to appear the contact part of the effective interaction $V_0$ must be anomalously small, i. e. the gas must be almost ideal. Providing that the condition of the type \eqref{nc} is compatible with the diluteness criterion $n_1 x_\ast\ll1$, the system condenses into a regular array of Bose-Einstein condensates.

At $T\ll T_c$ in the typical experimental conditions\cite{Butov2002} the condensates are dense and strongly-correlated. The physics of this state can be described by the phenomenological model introduced in Subsection D. It is presently unclear if such a state can exist as a true ground state of the gas (in the sense of a Bose-Einstein condensed metastable state), or it results from a dynamical equilibrium between the localized excitons and delocalized electron-hole plasma. Experiments on ultra-cold polar molecules in the bilayer geometry could enlighten this question. In particular, one could try to observe a transition from the weakly-interacting supersolid to the fragmented-condensate solid state first by going to the 2D cigar regime by increasing the particle density and then increasing the contact part of the two-body interaction.

In excitonic systems it should be possible to tune $V_0$ by exploiting the shape resonance in the pairwise interaction of excitons having opposite spins\cite{Andreev3, Andreev4}. The emergent physics is a $2D$ analog of the Feshbach resonance in atomic gases\cite{Feshbach}. Due to the Pauli exclusion of the electrons and holes, formation of the three-body and larger excitonic complexes is prohibited, which allows to explore the whole range of the 2D scattering lengths, including the unitary limit. At the many-body level, this should allow to observe a quantum phase transition from an exciton condensate to a superfluid of tightly-bound excitonic molecules (biexcitons). In the dilute regime, approaching the transition from the excitonic side results in reduction of $V_0$ and rotonization of the elementary excitation spectrum\cite{Andreev4}. 

The presence of a resonance in the two-body interaction of excitons thus may explain, how the system characterized by strong repulsion at $T\ll T_c$ can crystallize, when $T$ approaches $T_c$ from above. At $T\rightarrow T_c$ the chemical potential approaches the scattering threshold (zero energy in the thermodynamic limit). If the position of the resonance is sufficiently close to the threshold, the average effective interaction $V_0$ decreases due to resonant attraction and pairing of excitons. On the other hand, at $T\ll T_c$ the chemical potential is large, and the system is in the molecular phase, characterized by strong repulsive interaction between the dipolar molecules.

An intriguing prediction of our theory is that it may be possible to observe an abrupt extension of the coherence length over multiple periods of the MOES by reducing the exciton density and going to extremely low temperatures (below 0.1 K, see the discussion in the end of Subsection B, theory). As the system becomes more dilute, the fraction of the molecular superfluid in a bead decreases and contribution of the unpaired excitonic component to the average interaction becomes more important. As a result, one may expect to reduce the dimensionless interaction parameter $\eta$ down to the critical value \eqref{etac}. However, the bath temperature $T$ must be sufficiently low (on the order of $\mu$) in order to suppress the long-wave fluctuations of the phase - otherwise, one would merely destroy the crystal.\cite{Gorbunov} 

In conclusion, we have argued that the MOES may represent a form of the supersolid state of matter. Our theory identifies the MOES with a fragmented-condensate solid, akin to the fragmented BEC in an optical lattice\cite{Kasevich}. This state can be obtained from the coherent density wave by increasing repulsive interaction between the particles. The fragmented state is characterized by a different dependence of the wavelength $\lambda$ on the particle density $n$. In contrast to the coherent supersolid, the wavelength of the MOES increases with $n$. The obtained formula is confirmed by the recent experiments. We expect polar molecules in bilayer geometries to be particularly promising for further investigation of the subject.      

\section*{Aknowledgements}

The author thanks D. S. Petrov, G. V. Shlyapnikov and G. Pupillo for stimulating discussions. I am grateful to A. Bouzdine for a helpful critical reading of the manuscript. I also thank A. Cano for drawing my attention to Ref. [9]. The work was supported by the Government of the Russian Federation (Grant 074-U01) through ITMO Postdoctoral Fellowship scheme.

\section*{Appendix A: Topological transformation of the fragmented condensate conserving its interaction energy}

Let us calculate the volume $V$ of the following object in $(x,y,z)$ space:
\begin{equation}
\label{figure}
\frac{z}{h}=\left(\frac{x}{R_{x}}\right)^{2\alpha}+\left(\frac{y}{R_{y}}\right)^{2\alpha}+\beta\left(\frac{xy}{R_{x}R_{y}}\right)^{\alpha},
\end{equation}
where $(\alpha=1,\beta=0)$ or $(\alpha=2,\beta=2)$ and $z\leqslant h_0\leqslant h$. In the case $(\alpha=2,\beta=2)$ evaluation of $V$ formally corresponds to evaluation of the Thomas-Fermi energy of one condensate [see Eq. \eqref{TFenergy}]. The case $(\alpha=1,\beta=0)$ yields the number of particles. The height $h_0$ of the figure \eqref{figure} physically corresponds to the chemical potential $\mu$ (or, equivalently, the maximum of the density).

By introducing cylindrical coordinates $x=r\cos\theta$ and $y=r\sin\theta$, we write the area $S(z)$ of a cross-section at the height $z$ in the form
\begin{equation}
\begin{split}
S(z)=&\int\limits_{0}\limits^{2\pi}\int\limits_{0}\limits^{r(z)}rdrd\theta\\
=&\left(\frac{z}{h}\right)^{1/\alpha}R_{x}R_{y}\int\limits_{-\infty}\limits^{+\infty}(1+x^{2\alpha}+\beta x^{\alpha})^{-1/\alpha}dx
\end{split}
\end{equation}
so that the volume reads as
\begin{equation}
\begin{split}
V=&\int\limits_{0}\limits^{h_0}S(z)dz\\
=&\frac{R_{x}R_{y}h_0^{1/\alpha+1}}{(1/\alpha+1)h^{1/\alpha}}\int\limits_{-\infty}\limits^{+\infty}(1+x^{2\alpha}+\beta x^{\alpha})^{-1/\alpha}dx.
\end{split}
\end{equation}
One can see that the resulting expression is linear in $R_x$. Consequently, the volume of our object can be decomposed  into a sum of volumes $V_j$ of J similar objects (j=1,2,..., J) with the same height $h_0$, the same transverse diameter of the base ellipse $R_y$, and with the sum of $R_{x,j}$ being equal to $R_x$:
\begin{equation}
\label{partition}
V=V_{1}+V_{2}+...+V_{J}.
\end{equation}

\section*{Appendix B: Quantum pressure correction to the supersolid energy}

By introducing cylindrical coordinates at each unit cell we write Eq. \eqref{kin} in the form
\begin{equation}
E_\mathrm{kin}=E_\mathrm{kin}^{(\rho)}+E_\mathrm{kin}^{(\theta)},
\end{equation}
where
\begin{equation}
E_\mathrm{kin}^{(\rho)}=\frac{J\hbar^2}{2m}\int\limits_0^{2\pi}\int\limits_0^{+\infty}\left\vert\frac{\partial\Pi(\rho,\theta)}{\partial\rho}\right\vert^2 \rho d\rho d\theta
\end{equation}
and     
\begin{equation}
\label{angular}
E_\mathrm{kin}^{(\theta)}=\frac{J\hbar^2}{2m}\int\limits_0^{2\pi}\int\limits_0^{+\infty}\frac{1}{\rho^2}\left\vert\frac{\partial\Pi(\rho,\theta)}{\partial\theta}\right\vert^2 \rho d\rho d\theta
\end{equation}
with $J$ being the total number of unit cells and $\Pi(\rho,\theta)$ is the Thomas-Fermi profile of one bead (we omit the index $j$ for brevity). The radial integrals can be divided into two parts:
\begin{equation}
\label{twoparts}
\int\limits_0^{+\infty}d\rho=\int\limits_0^{R(\theta)-\varepsilon}d\rho+\int\limits_{R(\theta)-\varepsilon}^{+\infty}d\rho,
\end{equation}
where
\begin{equation}
R(\theta)=\frac{R_x}{\sqrt{\cos^2\theta+\sigma^2\sin^2\theta}}
\end{equation}
with
\begin{equation}
\sigma=\frac{R_x}{R_y}
\end{equation}
being the aspect ratio of a bead and the constant $\varepsilon$ is chosen such as $d(\theta)\ll\varepsilon\ll R(\theta)$. The characteristic thickness of the boundary $d(\theta)$ is given by
\begin{equation}
d(\theta)=\left(\frac{2m F(\theta)}{\hbar^2}\right)^{-1/3}
\end{equation}
with
\begin{equation}
\label{F}
F(\theta)=\sqrt{2\mu m\omega_x^2(\cos^2\theta+\sigma^2\sin^2\theta)}.
\end{equation}
By introducing $s=(\rho-R(\theta))/d(\theta)$ and performing the transformation
\begin{equation}
\Pi(\rho)=\sqrt{\frac{\hbar^2}{2m\tilde V_0 d^2}}\Phi(s)
\end{equation}
one can rewrite the Gross-Pitaevskii equation on $\Pi(\rho,\theta)$ close to $R(\theta)$ in the universal form\cite{Boundary}
\begin{equation}
\label{universal}
\phi''-(s+\phi^2)\phi=0.
\end{equation}
This form can be conveniently used to evaluate the second integral in \eqref{twoparts}, whose main contribution comes from the boundary region. Neglecting corrections vanishing as $\varepsilon/R(\theta)$ and integrating over $\theta$ one finds
\begin{equation}
E_\mathrm{kin}^{(\rho)}=J\frac{\pi}{2}\frac{\hbar^2}{m \tilde V_0}\mu\left(\ln\frac{R_x}{d_x}-\frac{5}{3}\ln\frac{\sigma+1}{2}-1+4C\right),
\end{equation}
where
\begin{equation*}
C=-\int\limits_{-\infty}^{+\infty}\ln(\sqrt{1+s^2}+s)\frac{d}{ds}[(\phi')^2\sqrt{1+s^2}]ds=0.176.
\end{equation*}
Expanding to the first order in powers of $(\sigma-1)$ yields
\begin{equation}
E_\mathrm{kin}^{(\rho)}=J\frac{\pi}{2}\frac{\hbar^2}{m \tilde V_0}\mu\left(\ln\frac{R_x}{d_x}-\frac{1}{6}+4C\right),
\end{equation}
where we have omitted the $J$-independent part. As regards the angular contribution \eqref{angular}, it can be shown to be proportional to $(\sigma-1)^2$ and thus can be neglected for our purposes. By using the relationship \eqref{eta} one can finally write the equation for the quantum pressure of the Thomas-Fermi supersolid in the form
\begin{equation}
E_\mathrm{kin}=J\frac{x}{\eta}k_B T_c^{0},
\end{equation}
where we have introduced
\begin{equation}
x=\frac{\pi^2}{12}\left(\ln\frac{R_x}{d_x}-\frac{1}{6}+4C\right).
\end{equation}
The dimensionless coefficient $x$ is defined by the topology of the supersolid density profile in the depleted regions in between the beads. For the harmonic potential model \eqref{F} the ratio $R_x/d_x$ scales as $\omega_x^{-1/3}$ and, therefore, depends on $J$. However, it is presently unclear to what extent the harmonic trap can be used to model the actual crystalline mean-field potential. For the sake of simplicity, it is reasonable to assume that the effective boundary width $d_x$ scales as the size of a unit cell $R_x$, which implies the parameter $x$ to be a constant (not dependent on $J$).


\begin{thebibliography}{99}

\bibitem{Gross}  E. P. Gross, Phys. Rev. \textbf{106}, 161 (1957); Annals of Physics \textbf{4}, 57 (1958).

\bibitem{Kirzhnits} D. A. Kirzhnits and Yu. A. Nepomnyashchii, Sov. Phys.
JETP \textbf{32}, 1191 (1971).

\bibitem{Nepomnyashchii} Yu. A. Nepomnyashchii, Theor. Math. Phys. \textbf{8}, 928 (1971).

\bibitem{Brazovskii} S. A. Brazovskii, Sov. Phys. JETP \textbf{41}, 85 (1975).

\bibitem{Pitaevskii} L. P. Pitaevskii, JETP Lett. \textbf{39}, 511 (1984).

\bibitem{Pomeau} Y. Pomeau and S. Rica, Phys. Rev. Lett. \textbf{72}, 2426 (1994).

\bibitem{Meystre} O. Dutta, R. Kanamoto, and P. Meystre, Phys. Rev. Lett. \textbf{99}, 110404 (2007).

\bibitem{Suto} A. Suto, J. Math. Phys. \textbf{50}, 032107 (2009).

\bibitem{Rica} S. Rica, Int. J. Bifurcation Chaos \textbf{19}, 2783 (2009).

\bibitem{Nozieres} P. Nozieres, J. Low Temp. Phys. \textbf{156}, 9 (2009).

\bibitem{Henkel} N. Henkel, R. Nath, and T. Pohl, Phys. Rev. Lett. \textbf{104}, 195302 (2010).

\bibitem{Cinti} F. Cinti, P. Jain, M. Boninsegni, A. Micheli, P. Zoller, and G. Pupillo, Phys. Rev. Lett. \textbf{105}, 135301 (2010).

\bibitem{Li} X. Li, W. V. Liu and C. Lin, Phys. Rev. A \textbf{83}, 021602 (2011).

\bibitem{Saccani} S. Saccani, S. Moroni, and M. Boninsegni, Phys. Rev. B
\textbf{83}, 092506 (2011).

\bibitem{Chen} Y. Chen, J. Ye and G. Tian, J. Low Temp. Phys. \textbf{169}, 149 (2012).

\bibitem{Josserand} P. Mason and C. Josserand, Phys. Rev. B \textbf{88}, 224506 (2013).

\bibitem{DiluteSupersolid} Zhen-Kai Lu, Yun Li, D. S. Petrov, and G. V. Shlyapnikov, Phys. Rev. Lett. \textbf{115}, 075303 (2015);

\bibitem{Molecules} S. Inouye \textit{et al.}, Phys. Rev. Lett. \textbf{93}, 183201 (2004); S. Ospelkaus \textit{et al.}, Nat. Phys. \textbf{4}, 622 (2008); M. H. G. de Miranda et al., Nat. Phys. \textbf{7}, 502 (2011); Tetsu Takekoshi et al., Phys. Rev. Lett. \textbf{113}, 205301 (2014); Mingyang Guo et al., Phys. Rev. Lett. \textbf{116}, 205303 (2016).

\bibitem{Santos} L. Santos, G. V. Shlyapnikov, and M. Lewenstein, Phys. Rev. Lett. \textbf{90}, 250403 (2003).

\bibitem{Andrews} M. R. Andrews et al., Science \textbf{275}, 637 (1997).

\bibitem{Ketterle} J. Stenger et al., Phys. Rev. Lett. \textbf{82}, 4569 (1999).

\bibitem{Walraven} I. Shvarchuck et al., Phys. Rev. Lett. \textbf{89}, 270404 (2002).

\bibitem{ColdExcitons} L. V. Butov, Journal of Physics: Condensed Matter \textbf{19}, 295202 (2007);  L. V. Butov, JETP 122, 434 (2016).

\bibitem{dense} R. Rapaport, G. Chen, Journal of Physics: Condensed Matter \textbf{19}, 295207 (2007).

\bibitem{Ivchenko} E. L. Ivchenko, \textit{Optical Spectroscopy of Semiconductor Nanostructures} (Alpha Science International, Harrow, UK, 2005).

\bibitem{Bieker} S. Bieker, T. Henn, T. Kiessling, W. Ossau, and L. W. Molenkamp, Phys. Rev. Lett. \textbf{114}, 227402 (2015).

\bibitem{Kavokin} A. A. High, A. T. Hammack, J. R. Leonard, Sen Yang, L. V. Butov, T. Ostatnicky, M. Vladimirova, A. V. Kavokin, T. C. H. Liew, K. L. Campman, and A. C. Gossard, Phys. Rev. Lett. \textbf{110}, 246403 (2013).

\bibitem{Pfeiffer} Yongbao Sun, Patrick Wen, Yoseob Yoon, Gangqiang Liu, Mark Steger, Loren N. Pfeiffer, Ken West, David W. Snoke, and Keith A. Nelson, Phys. Rev. Lett. \textbf{118}, 016602 (2017).

\bibitem{Butov2002} L.~V.~Butov, A.~C.~Gossard, and D.~S.~Chemla, Nature
(London) \textbf{418}, 751 (2002).

\bibitem{Repulsive} Sen Yang, A.~V. Mintsev, A.~T.~Hammack, L.~V.~Butov, and
A.~C.~Gossard, Phys. Rev. B \textbf{75}, 033311 (2007).

\bibitem{High2012} A.~A.~High, J.~R.~Leonard, A.~T.~Hammack, M.~M.~Fogler,
L.~V.~Butov, A.~V.~Kavokin, K.~L.~Campman, A.~C.~Gossard, Nature (London) 
\textbf{483}, 584 (2012).

\bibitem{Pittsburgh} D. Snoke, S. Denev, Y. Liu, L. Pfeiffer and K. West, Nature (London) \textbf{418}, 754 (2002).

\bibitem{Yang} S. R. E. Yang, Q. H. Park, and J. Yeo, Int. J. Mod. Phys. B \textbf{18}, 3797 (2004).

\bibitem{Paraskevov} A. V. Paraskevov and T. V. Khabarova, Phys. Lett. A \textbf{368}, 151 (2007).

\bibitem{Levitov} L. S. Levitov, B. D. Simons and L. V. Butov, Phys. Rev. Lett. \textbf{94}, 176404 (2005).

\bibitem{Sugakov} A. A. Chernyuk and V. I. Sugakov, Phys. Rev. B \textbf{74}, 085303 (2006).

\bibitem{Liu} C. S. Liu, H. G. Luo, W. C. Wu, J. Phys.: Condens. Matter \textbf{18}, 9659 (2006).

\bibitem{Wilkes} J. Wilkes, E. A. Muljarov, and A. L. Ivanov, Phys. Rev. Lett. \textbf{109}, 187402 (2012).

\bibitem{Andreev1} S. V. Andreev, Phys. Rev. Lett. \textbf{110}, 146401 (2013).

\bibitem{Andreev2} S. V. Andreev, A. A. Varlamov and A. V. Kavokin, Phys. Rev. Lett. \textbf{112}, 036401 (2014).

\bibitem{Andreev3} S. V. Andreev, Phys. Rev. B \textbf{92}, 041117(R) (2015).

\bibitem{Andreev4} S. V. Andreev, Phys. Rev. B \textbf{94}, 140501(R) (2016).

\bibitem{Commensurability} Sen Yang, L. V. Butov, B. D. Simons, K. L. Campman and A. C. Gossard, Phys. Rev. B \textbf{91}, 245302 (2015).

\bibitem{Kasevich} C. Orzel, A. K. Tuchman, M. L. Fenselau, M. Yasuda, and M. A. Kasevich, Science \textbf{291}, 2386 (2001).

\bibitem{Stringari} L. P. Pitaevskii and S. Stringari, \textit{Bose-Einstein condensation and superfluidity} (Oxford University Press, Oxford, 2016).

\bibitem{Haldane} F. D. M. Haldane, Phys. Rev. Lett. \textbf{47}, 1840 (1981).

\bibitem{Popov} Rigorous proof of the validity of the Bogoliubov approach in the absence of a true condensate in a 1D weakly-interacting gas can be found in [V. N. Popov, \textit{Functional Integrals in Quantum Field Theory and Statistical Physics} (Reidel, Dordrecht, 1983)].

\bibitem{Tonks} The transition to a Tonks gas (or, to be more precise, to a super-Tonks gas) should be discussed with care. In fact, when the 2D scattering length $a$ becomes larger $x_\ast$ and eventually approaches $a_y$, a confinement-indused resonance (CIR) may come into play [M. Olshanii, Phys. Rev. Lett. \textbf{81}, 938 (1998)]. The effect of CIR on the behavior of dipoles in a quasi-1D tube was studied in [S. Sinha and L. Santos, Phys. Rev. Lett. \textbf{99}, 140406 (2007)].

\bibitem{Collapse} S. Komineas and N. R. Cooper, Phys. Rev. A \textbf{75}, 023623 (2007); N. G. Parker, C. Ticknor, A. M. Martin, and D. H. J. O'Dell, Phys. Rev. A \textbf{79}, 013617 (2009).

\bibitem{2Ddipoles} J. Levinsen, N. R. Cooper and G. V. Shlyapnikov, Phys. Rev. A \textbf{84}, 013603 (2011).

\bibitem{Thesis} S. V. Andreev, \textit{Bose-Einstein condensation of excitons in semiconductor nanostructures}, PhD at the University of Montpellier (2014). 

\bibitem{Landau} L. D. Landau and E. M. Lifshitz, \textit{Statistical Physics, Part 1}
(Pergamon Press, New York, 1969).

\bibitem{Petrov} D. S. Petrov, Phys. Rev. Lett. \textbf{112}, 103201 (2014).

\bibitem{Subwavelength} A. Gonzalez-Tudela, C.-L. Hung, D. E. Chang, J. I. Cirac and H. J. Kimble, Nature Photon. \textbf{9}, 320 (2015).

\bibitem{Formation} L.V. Butov, L.S. Levitov, A.V. Mintsev, B.D. Simons, A.C. Gossard, and D.S. Chemla, Phys. Rev. Lett. \textbf{92}, 117404 (2004).

\bibitem{Andreev5} S. V. Andreev, Phys. Rev. B \textbf{94}, 165308 (2016).

\bibitem{BCSmodel} In ultra-cold atoms, the model of a resonantly paired 2D binary mixture of dipolar Bosons can be implemented in the system of polar molecules in the bilayer geometry \textit{without} tunneling. An analogy of the exciton spin here would be the layer index. The interaction potential for two molecules in different layers can be shown to admit a bound state [M. Klawunn, A. Pikovski, and L. Santos, Phys. Rev. A \textbf{82}, 044701 (2010)]. More about the low-energy scattering properties of such systems can be found in [M. A. Baranov, A. Micheli, S. Ronen, and P. Zoller, Phys. Rev. A  \textbf{83}, 043602 (2011)]. One may expect that, under appropriate conditions, this bound state can become a resonance. It is important to stress, that the effective reduction of $V_0$, the predicted roton instability and the transition to a supersolid phase here should be searched on the attractive side of the resonance and not in the vicinity of a weakly-bound state. In the latter case the contact interaction vanishes only in the channel where the two molecules belong to different layers, whereas it remains finite (repulsive) for the intralayer scattering.

\bibitem{Boundary} F. Dalfovo, L. P. Pitaevskii and S. Stringari, Phys. Rev. A \textbf{54}, 4213 (1996).

\bibitem{Lattices} B. P. Anderson and M. Kasevich, Science \textbf{282}, 1686 (1998).

\bibitem{Josephson} B. D. Josephson, Phys. Lett. \textbf{1}, 251 (1962).

\bibitem{Fluctuations} L. P. Pitaevskii and S. Stringari, Phys. Rev. Lett. \textbf{87}, 180402 (2001).

\bibitem{Carruthers} P. Carruthers and M. M. Nieto, Rev. Mod. Phys. \textbf{40}, 411 (1968).

\bibitem{Leggett} A. J. Leggett, Rev. Mod. Phys. \textbf{73}, 307 (2001).

\bibitem{Bradley} R. M. Bradley and S. Doniach, Phys. Rev. B \textbf{30}, 1138 (1984).

\bibitem{Rontani} M. Rontani and L. J. Sham, Phys. Rev. B \textbf{80}, 075309 (2009).

\bibitem{Zapata} I. Zapata, F. Sols and A. J. Leggett, Phys. Rev. A \textbf{57}, R28 (1998).

\bibitem{zimmermann} C. Schindler and R. Zimmermann, Phys. Rev. B \textbf{78}, 045313 (2008).

\bibitem{Combescot} M. Combescot, R. Combescot, and F. Dubin, Rep. Prog. Phys.
80, 066501 (2017).

\bibitem{BellLabs} R. Rapaport, Gang Chen, D. Snoke, Steven H. Simon, Loren
Pfeiffer, Ken West, Y. Liu and S. Denev, Phys. Rev. Lett. \textbf{92}, 117405 (2004).

\bibitem{Thermalization} L. V. Butov, A. L. Ivanov, A. Imamoglu, P. B. Littlewood, A. A. Shashkin, V. T. Dolgopolov, K. L. Campman, and A. C. Gossard, Phys. Rev. Lett. \textbf{86}, 5608 (2001).

\bibitem{footnote} The expression for the critical temperature of the fragmented-condensate solid accounting for repulsive interactions between the particles reads\cite{Andreev2}
\begin{equation*}
\label{tc}
T_{c}=\frac{T_{c}^{0}}{\sqrt{1+\frac{x^2(\eta)}{4\eta^{2}}}},
\end{equation*}
where $x(\eta)$ is a root of $\pi^{2}x=12\eta^{2}g_{1}(e^{-x})$ and $g_{1}(x)$ is the Bose function. The value of the contact interaction parameter $\eta$ here should be taken in the critical region $T\rightarrow T_c$, and it is expected to be vanishingly small (see discussion in the main text). 

\bibitem{Feshbach} C. Chin, R. Grimm, P. Julienne, and E. Tiesinga, Rev. Mod. Phys. \textbf{82}, 1225 (2010).

\bibitem{Gorbunov} An alternative way to suppress the long wave fluctuations of the phase is to reduce the longitudinal size of the system. In this regard, it is worth to mention the experimental work [A. V. Gorbunov and V. B. Timofeev, JETP Lett. \textbf{83}, 146 (2006)], where a regular pattern of excitonic clusters was observed in a ring-shaped electrostatic trap of a 5-$\mu$m diameter. The wavelength of the structure was found to \textit{decrease} when increasing the exciton density.    


\end{thebibliography}
\end{document}